\documentclass{article} % For LaTeX2e
\usepackage{iclr2027_conference,times}

\usepackage{amsmath,amsfonts,bm}

\def\eqref#1{equation~\ref{#1}}
\def\1{\bm{1}}

\DeclareMathAlphabet{\mathsfit}{\encodingdefault}{\sfdefault}{m}{sl}
\SetMathAlphabet{\mathsfit}{bold}{\encodingdefault}{\sfdefault}{bx}{n}

\usepackage{hyperref}
\usepackage{url}
\usepackage{graphicx}
\usepackage{amssymb}
\usepackage{enumitem}
\usepackage{booktabs}
\usepackage{tabularx}
\usepackage{makecell}
\usepackage{subcaption}
\usepackage{siunitx}
\usepackage{float}
\usepackage{array}
\usepackage{booktabs}
\usepackage{listings}

\lstdefinestyle{ciappendix}{
    basicstyle=\ttfamily\footnotesize,
    columns=fullflexible,
    keepspaces=true,
    showstringspaces=false,
    breaklines=true,
    breakatwhitespace=false,
    frame=single,
    framerule=0.3pt,
    framesep=5pt,
    numbers=none,
    captionpos=b,
    aboveskip=0.7em,
    belowskip=0.7em,
    tabsize=2,
    mathescape=false,
    language={}
}

\title{Judging Before Propagation: Task-Graph Contextual Integrity for Multi-Agent Privacy}

\author{Hengyu Wu, Blaise Delattre, Yang Cao \\
Department of Computer Science, School of Computing \\
Institute of Science Tokyo \\
\texttt{wu.h.bc16@m.isct.ac.jp}, \texttt{delattre.b.aa@m.titech.ac.jp}, \texttt{cao@c.titech.ac.jp} \\
}

\iclrfinalcopy % Show author names for arXiv preprint

\begin{document}

\maketitle

% Override the camera-ready publication notice
\fancyhead[L]{Preprint}

\begin{abstract}
Multi-agent LLM systems increasingly rely on persistent memory and inter-agent communication. These capabilities create privacy risks that extend beyond individual interactions: a release appropriate in isolation may become inappropriate when combined with information previously disclosed to a recipient or propagated to downstream agents along the planned workflow. Existing Contextual Integrity (CI) mechanisms for agentic systems typically rely on local judgments that assess each disclosure in isolation, with or without regard to the recipient's history, but do not account for how information propagates downstream. However, compliance at the current boundary does not guarantee appropriateness across the workflow. We introduce \textbf{Task-Graph Contextual Integrity (TG-CI)}, to our knowledge the first task-graph-native formalization of Contextual Integrity for multi-agent systems. TG-CI represents a workflow as a \textbf{Multi-Agent Task Graph (MATG)} and extends CI from atomic transfers to stateful, graph-level information flows by jointly considering an item's intended propagation scope and recipients' evolving disclosure states. We operationalize TG-CI through \emph{propagation-scoped authorization} under two complementary policies: \textit{Full-Scope}, which requires admissibility across the complete downstream scope, and \textit{Auto-Region}, which confines each item to its maximal admissible region using lightweight frontier gates. We further introduce \textbf{MAPLE-Bench}, a benchmark of 144 cases across four scenarios that jointly vary MATG topology, disclosure history, compromised-agent placement, and item-level privacy policies. In our primary experimental configuration, TG-CI reduces the mean unauthorized disclosure count by \(85.8\text{--}94.1\%\) relative to the Local-History CI baseline, while retaining \(93.6\text{--}96.1\%\) of the baseline's task accuracy, with the two enforcement policies providing complementary privacy--cost--utility trade-offs.
\end{abstract}

\section{Introduction}
\begin{figure}[htbp]
    \centering
    \includegraphics[width=\linewidth]{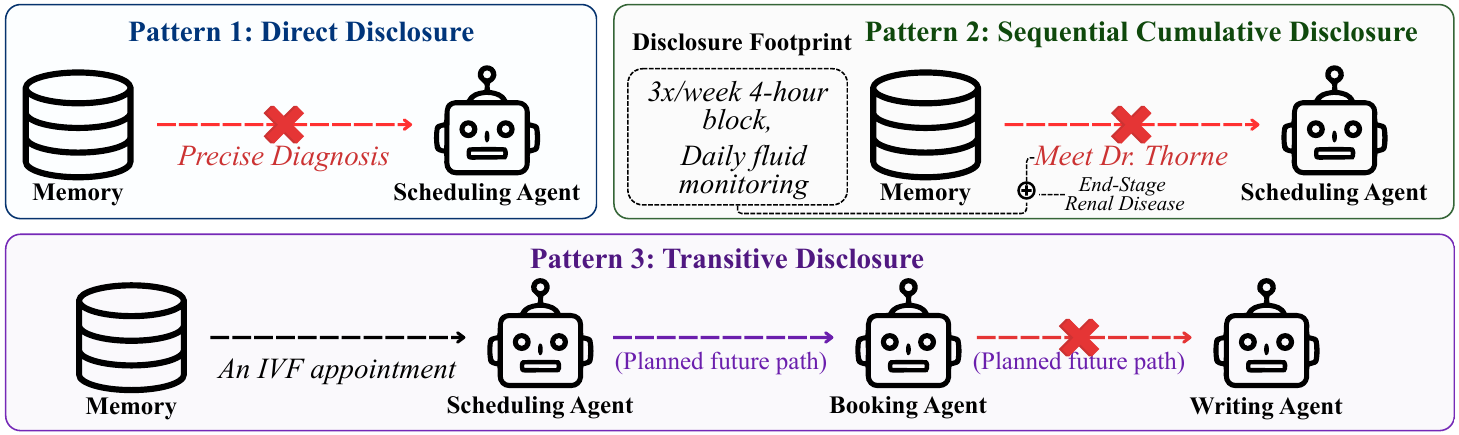}
    \vspace{-5mm}
    \caption{Three failure patterns in multi-agent systems. \textbf{Direct disclosure} violations occur when the current transfer itself is contextually inappropriate. \textbf{Sequential Cumulative} disclosure violations occur when a current disclosure combined with previous disclosures could reveal inappropriate sensitive information. \textbf{Transitive disclosure} violation occurs when an initially appropriate disclosure would propagate to an inappropriate downstream recipient along the foreseeable task workflow.}
    \label{fig:fail_pattern}
    \vspace{-2mm}
\end{figure}

Multi-agent systems increasingly combine persistent memory with task delegation and inter-agent communication to solve complex problems \citep{ijcai2024Guo, arxiv2025tran}. While these capabilities improve coordination and continuity, they also enlarge the privacy attack surface: sensitive information retrieved by one agent can be retained \citep{arxiv2026Lahjouji}, combined with prior knowledge \citep{arxiv2025patil}, and propagated to other agents throughout the workflow \citep{arxiv2026Lahjouji,fcomp2026Bhosale}. Figure~\ref{fig:fail_pattern} illustrates three primary privacy violation modes in such systems. This risk is especially concerning when some agents are compromised.
%or otherwise contextually inappropriate recipients. 
For instance, while disclosing medical appointment metadata to a scheduling service may be permissible in isolation, aggregating this with preexisting recipient data might expose sensitive clinical details, which may then propagate to a compromised downstream notification agent. In such settings, protecting an individual memory access in isolation is insufficient, because the privacy consequences of a disclosure depend also on prior disclosure and the downstream multi-agent workflow.
%not only on who receives it now, but also

\begin{table}[htbp]
    \centering
    \small
    \label{tab:related_work}
    \begin{tabular}{lcccc}
        \toprule
        \textbf{Paper} & \textbf{CI Grounded} & \makecell{\textbf{Direct}\\\textbf{Disclosure}} & \makecell{\textbf{Cumulative}\\\textbf{Disclosure}} & \makecell{\textbf{Transitive}\\\textbf{Disclosure}} \\
        \midrule
        AirGapAgent \citep{acm2024Bagdasarian}& Y & Y & N & N \\
        The Sum Leaks \citep{arxiv2025patil}    & N & N & Y & N \\
        OCELOT \citep{arxiv2026xie}           & Y & Y & Y & N \\
        Minim \citep{icml2026yu}            & Y & Y & N & N \\
        CAMP \citep{arxiv2026Panjwani}             & N & Y & Y & N \\
        \midrule
        \textbf{\emph{TG-CI (Ours)}}         & \textbf{Y} & \textbf{Y} & \textbf{Y} & \textbf{Y} \\
        \bottomrule
    \end{tabular}
    \caption{Comparison of related work on multi-agent privacy \textit{evaluation and} protection.}
    \vspace{-6mm}
\end{table}

Contextual Integrity (CI) provides a natural foundation for privacy in multi-agent workflows \citep{springer2025chen}, yet how downstream task structure should inform history-aware disclosure judgments remains underexplored.
CI evaluates whether an information flow is appropriate within its surrounding context \citep{nissenbaum2004privacy,sp2006barth}, rather than relying solely on fixed sensitivity labels.
%as in conventional Personally Identifiable Information (PII) masking \cite{amia1996Sweeney}
However, existing privacy mechanisms for agentic systems largely operationalize privacy at individual disclosure or interaction boundaries, determining whether a current transfer is acceptable based on its immediate context \citep{acm2024Bagdasarian,icml2026yu}. The AgentSCOPE \citep{arXiv2026Ngong} benchmark further expands it to multi-tool agentic workflow by labeling internal channels but still considers each information flow independently and provides no defense. Recent work has begun to address cumulative disclosure risks through trajectory-level privacy budgeting \citep{arxiv2026xie}. Meanwhile, empirical studies identify inter-agent communication as a major channel of privacy leakage in multi-agent systems \citep{access2026Yagoubi}, yet comparatively little work considers how the downstream task structure itself should influence an earlier disclosure decision. This leaves a central question underexplored: \emph{how can we extend contextual integrity to evaluate an information flow by jointly accounting for direct, cumulative, and transitive disclosure risks within a multi-agent task structure?}

To address this gap, we introduce \textbf{Task-Graph Contextual Integrity (TG-CI)}, operationalize it through propagation-scoped authorization with dual-mode enforcement, and develop \textbf{MAPLE-Bench} for controlled evaluation of multi-agent privacy.
TG-CI uses a \textbf{Multi-Agent Task Graph (MATG)} to evaluate a current information flow in light of recipients' disclosure histories and intended downstream propagation, jointly accounting for direct, cumulative, and transitive disclosure risks.
To make these judgments effective beyond the release boundary, the runtime binds each released item to an authorized propagation scope, allowing either conservative admission or selective downstream containment.
MAPLE-Bench jointly varies task structure, recipient disclosure history, and compromised-agent placement to assess whether these controls limit all three disclosure risks while preserving task utility.

In summary, our contributions are:
\begin{enumerate}[leftmargin=*, nosep, label=(\arabic*)]
    \item \textbf{Task-graph-native contextual integrity.}
    We formalize TG-CI, to our knowledge the first task-graph-native CI formalism that judges a current disclosure jointly with its intended downstream propagation scope and recipients' evolving disclosure states. TG-CI preserves classical CI semantics at each atomic transfer while lifting the judgment to a graph-level information flow.

    \item \textbf{Propagation-scoped authorization and dual-mode enforcement.}
    We operationalize each positive TG-CI decision as a propagation-scoped authorization and introduce two complementary enforcement policies. Full-Scope Mode blocks an item if any transfer in its complete downstream closure is inadmissible, whereas Auto-Region Authorization computes the maximal rooted admissible region and removes the item at its propagation frontier. This design avoids expensive and repeated contextual re-evaluation along downstream routes while offering complementary trade-offs among privacy protection, computational cost, and task utility.

    \item \textbf{A benchmark for multi-agent privacy.} We release \textbf{MAPLE-Bench} (Multi-Agent Privacy Leakage Evaluation Benchmark), a reusable benchmark of 144 cases spanning four application scenarios, diverse MATG topologies, and compromised-agent configurations. Each case provides item-level memory, privacy policies and ground-truth annotations for
    %direct, cumulative, and transitive 
    disclosure violations, enabling controlled evaluation of information exposure and propagation containment.

    %\item \textbf{Comprehensive empirical evaluation. (Depend on space)}
    %We evaluate TG-CI under compromised-agent attacks across diverse multi-agent tasks, task-graph structures, and sensitive-memory scenarios, comparing against unprotected and local-flow privacy defenses in terms of information leakage, adversarial exposure, task utility, propagation containment, and runtime overhead.
\end{enumerate}

\section{Problem Setting}
\label{sec:ps}
\paragraph{System model.}
We consider a multi-agent system consisting of a set of agents $V$ that collaboratively execute a user task using persistent memory and inter-agent communication. At execution, we represent the current workflow as a directed, labeled \textbf{Multi-Agent Task Graph (MATG)} 
\begin{equation}
    G=(V,E,\tau,\pi),
    \label{eq:task-graph}
\end{equation}
where $E \subseteq V \times V$ contains planned task-dependent delegation or information-transfer relations, $\tau(v)$ denotes the current subtask assigned to agent $v$, and $\pi(e)$ denotes the edge transmission rule associated with edge $e$. The MATG differs from a communication-connectivity graph: $(u, v) \in E$ indicates that the current workflow anticipates a task-relevant interaction from $u$ to $v$, not merely that the two agents are technically able to communicate.

% \paragraph{Memory model.}
% Agents have access to a persistent memory system $\mathcal{M}$ that stores information accumulated from users or prior interactions. We model a memory record as a set of semantically meaningful information items $m_i=\langle d_i,a_i,x_i\rangle$, where $d_i$ denotes the data subject, $a_i$ the information attribute or type, and $x_i$ its value or content. During task execution, an agent $v\in V$ may issue a memory query $q_v$ and receive a retrieved subset $\mathcal{R}(q_v)\subseteq\mathcal{M}$. 
% %Retrieved information becomes available to the requesting agent and may subsequently influence messages or outputs passed along the MATG. 
% Our formulation is agnostic to the underlying memory architecture: $\mathcal{M}$ may represent shared, agent-specific, or hierarchical memory, provided that memory access occur through a system-mediated interface.

\paragraph{Memory model.}
Agents access a persistent memory system $\mathcal{M}$ that stores information accumulated from
%users or prior 
interactions. We model a memory record as a set of information items $m_i=\langle d_i,a_i,x_i\rangle$, where $d_i$ is the data subject, $a_i$ the information attribute or type, and $x_i$ its value or content. This granularity matches the subject and attribute parameters of CI \citep{sp2006barth} and the field-level user data in prior CI evaluations of agents \citep{acm2024Bagdasarian,bench_privacylens}. During task execution, an agent $v\in V$ may issue a memory query $q_v$ and receive a retrieved subset $\mathcal{R}(q_v)\subseteq\mathcal{M}$. The formulation does not depend on the memory architecture: $\mathcal{M}$ may be shared, agent-specific, or hierarchical, as in current agent memory systems \citep{packer2024memgptllmsoperatingsystems,chhikara2025mem0,zhang2025gmemory}. We only require that every memory access goes through a system-mediated interface. Access-controlled shared memory makes the same requirement \citep{rezazadeh2025collaborativememorymultiusermemory}.

% \paragraph{Adversary model.}
% We assume that an unknown subset $C \subseteq V$, with $|C| \le f$, may be compromised before execution. A compromised agent may deviate arbitrarily from its intended behavior: it can issue overbroad or adversarial memory requests, misrepresent task necessity, and transform sensitive content through unauthorized inter-agent messages. The adversary observes all information delivered to agents in $C$ and seeks to maximize the sensitive information learned by the compromised coalition and maximize inappropriate information flow within the system. However, we assume adversarial communication occurs through system-mediated memory and messaging interfaces; out-of-band channels are outside our threat model. [WRITE IN DEFENSIVE WAY, IN END OF IT]

\paragraph{Adversary model.}
An unknown subset $C\subseteq V$ with $|C|\le f$ may be compromised before execution, as in prior studies of malicious agents in LLM multi-agent systems \citep{huang2025resilience, bench_mama}. A compromised agent may deviate arbitrarily from its intended behavior: it can issue overbroad or adversarial memory requests, misrepresent task necessity, and transform sensitive content in inter-agent messages. The adversary observes all information delivered to agents in $C$ and seeks to maximize the protected information learned by the coalition and the inappropriate flows in the system. We assume that adversarial communication goes through the system-mediated memory and messaging interfaces, as in information-flow control for agents \citep{debenedetti2025camel,costa2025fides}. Out-of-band channels and covert encodings in permitted messages, such as steganographic collusion \citep{motwani2024secret}, are outside our threat model.

\paragraph{Privacy objective.}
Let $D_t(v)$ summarize the privacy-relevant information delivered to
recipient $v$ through monitored channels before execution step $t$
(formalized in Section~\ref{sec:tgci}).
Given the MATG $G$ and disclosure states $D_t$, we seek a CI-based
judgment of whether an item $m_i$ may flow from a source
$s\in V\cup\{\mathcal{M}\}$ to a recipient $r_0\in V$ under an
intended propagation scope $P$:
\[
    J(m_i,s,r_0,P \mid G,D_t)\in\{0,1\},
\]
where $J=1$ denotes admissibility.
An admissible flow must not cause direct, cumulative, or transitive disclosure violations within its intended propagation scope $P$.
%An admissible flow must be appropriate at the current transfer (\emph{direct disclosure}), reveal no prohibited information through combination with recipients' disclosure states (\emph{cumulative disclosure}), and remain appropriate along its intended downstream transfers in $P$ (\emph{transitive disclosure}).
The objective is to prevent inappropriate exposure and propagation while supporting task-relevant information use, not to ensure secrecy from compromised agents when access is contextually justified.

\section{Task-Graph Contextual Integrity}
\label{sec:tgci}

%In a multi-agent system, a disclosure to one agent is often only the first step of a planned sequence of downstream agent interactions. Moreover, whether a new disclosure is appropriate may depend on information that a downstream recipient has already received. 
TG-CI lifts an atomic CI judgment to a \emph{task-graph
information flow}: a protected information item is judged together
with its intended propagation scope and the evolving disclosure state
of the recipients along that scope.

\paragraph{Recipient disclosure state.}
To capture cumulative disclosure risks, TG-CI maintains the
recipient-specific disclosure state introduced in
Section~\ref{sec:ps} as a trusted mapping
\begin{equation}
    D_t:V\rightarrow\mathcal{H},
\end{equation}
where $\mathcal{H}$ denotes the space of privacy-relevant disclosure
summaries and $D_t(v)$ records the information delivered to recipient
$v$ through monitored channels before execution step $t$.
Thus, $D_t(v)$ represents the system-observable disclosure footprint
of $v$.

If $\Delta_t(v)$ denotes the privacy-relevant information actually
delivered to $v$ at step $t$, after any generalization or communication
filtering, the state evolves according to
\begin{equation}
    D_{t+1}(v)
    =
    D_t(v)\oplus\Delta_t(v),
    \label{eq:disclosure-update}
\end{equation}
where $\oplus$ incorporates the new disclosure into the recipient's
existing state. Authorization alone does not update $D_t(v)$; the
state changes only when information is actually delivered.

Given the MATG $G$, the contextual state under which TG-CI evaluates
a proposed information flow is
\begin{equation}
    \mathcal{S}_t
    =
    \langle G,D_t\rangle.
    \label{eq:tgci-state}
\end{equation}

\paragraph{Task-Graph Information Flow.}
%For a data subject $d$ and disclosed information attribute $a$,
% Let $s$ denote the information source, which may be an agent or the
% memory system $\mathcal{M}$. TG-CI distinguishes the overall MATG
% $G=(V,E,\tau,\pi)$ from the intended propagation of $a$, represented
% as a rooted, labeled propagation graph
Let $s$ denote the information source. The source is an agent or the
memory system $\mathcal{M}$. TG-CI distinguishes the MATG
$G=(V,E,\tau,\pi)$ from the propagation scope of a disclosure,
denoted $P$. The propagation scope is a rooted, labeled graph
\begin{equation}
    P=(V^P,E^P,\tau^P,\pi^P),
    \qquad
    \operatorname{root}(P)=s.
    \label{eq:propagation-scope}
\end{equation}
Here, $V^P\subseteq V\cup\{s\}$ and $P$ includes the initial
disclosure edge $(s,r_0)$ to the immediate recipient $r_0\in V$.
If $s\in V$, this edge must belong to the MATG; if
$s=\mathcal{M}\notin V$, it is added as a memory-release edge.
All subsequent edges satisfy $E^P\setminus\{(s,r_0)\}\subseteq E$,
and every node in $V^P\setminus\{s\}$ is reachable from $s$ through
a directed path in $P$.

%Let $e_0=(s,r_0)$ denote the initial disclosure edge to the immediate
%%recipient $r_0\in V$. We require
%\begin{equation}
%    V^P\subseteq V\cup\{s\},
%    \qquad
%    e_0\in E^P,
%    \qquad
%    E^P\setminus\{e_0\}\subseteq E,
%\end{equation}
%and every node in $V^P\setminus\{s\}$ must be reachable from $s$
%through a directed path in $P$. If $s\in V$, then $e_0\in E$; if
%$s=\mathcal{M}\notin V$, $e_0$ is added as a memory-release edge.

The node-task annotations in $P$ are inherited from the MATG, such
that $\tau^P(v)=\tau(v)$ for every $v\in V^P\cap V$.
%:
%\begin{equation}
%    \tau^P(v)=\tau(v),
%    \qquad
%    v\in V^P\cap V.
%\end{equation}
Similarly, the edge transmission rule of each propagation edge is
defined as
\begin{equation}
    \pi^P(e)=
    \begin{cases}
        \pi(e), & e\in E^P\cap E,\\
        \pi_0,  & e=e_0\notin E,
    \end{cases}
    \label{eq:propagation-rule}
\end{equation}
where $\pi_0$ is the edge transmission rule governing the release
from memory to $r_0$.

A \emph{task-graph information flow} is then defined as
\begin{equation}
    \Phi=\langle d,a,P\rangle.
    \label{eq:task-graph-flow}
\end{equation}

\paragraph{Atomic CI flow.}
TG-CI preserves the standard five-parameter CI representation at each
individual information-transfer boundary. For every propagation edge
$e=(u,v)\in E^P$, we instantiate a task-conditioned transmission
principle
\begin{equation}
    \rho^P(e)
    =
    \operatorname{Inst}
    \left(
        \pi^P(e),\tau^P(v)
    \right),
    \label{eq:task-conditioned-tp}
\end{equation}

The corresponding atomic CI flow is
\begin{equation}
    \phi_e(a)
    =
    \left\langle
        d,\,
        u,\,
        v,\,
        a,\,
        \rho^P(e)
    \right\rangle.
    \label{eq:atomic-ci-flow}
\end{equation}

TG-CI evaluates this flow against the recipient's evolving disclosure
state. We define the contextual admissibility of edge $e$ as
\begin{equation}
    \mathsf{Adm}_t(e,a)
    =
    \mathsf{CI}
    \left(
        \phi_e(a)
        \mid
        D_t(v)
    \right).
    \label{eq:edge-admissibility}
\end{equation}

\begin{figure}[htbp]
    \centering
    \includegraphics[width=\linewidth]{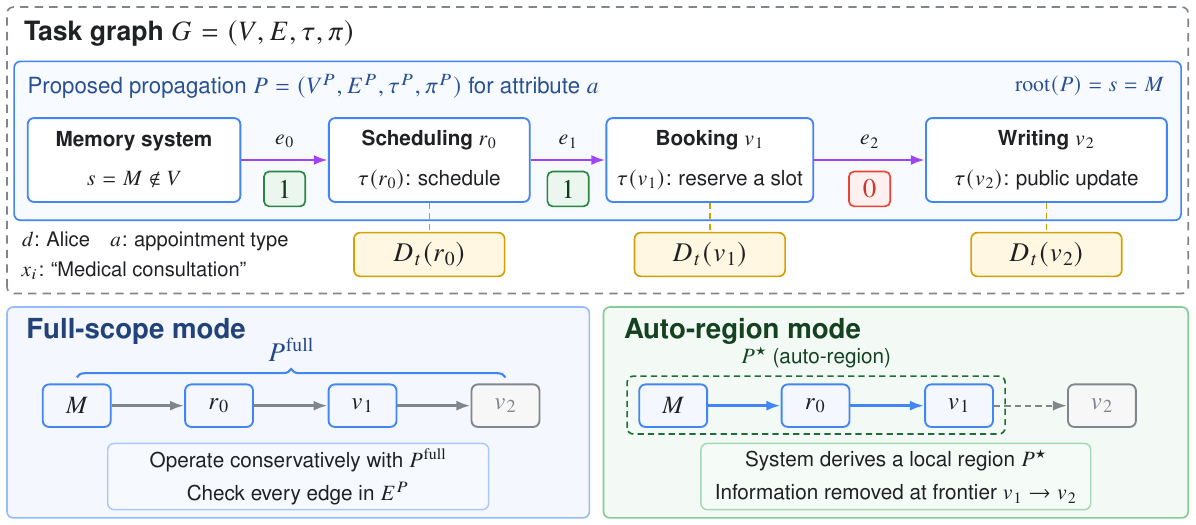}
    \vspace{-5mm}
    \caption{Overview of the operationalization of TG-CI and two enforcement policies.}
    \label{fig:workflow}
    \vspace{-2mm}
\end{figure}

\paragraph{Task-Graph Contextual Integrity.}
TG-CI treats $\Phi=\langle d,a,P\rangle$ as a single
task-graph-level information flow and judges it under the contextual
state $\mathcal{S}_t=\langle G,D_t\rangle$. The proposed flow is
contextually appropriate if every atomic transfer contained in its
propagation scope $P$ is contextually admissible:
\begin{equation}
\begin{aligned}
    \mathsf{TGCI}
    \left(
        \Phi,\mathcal{S}_t
    \right)
    &=
    \bigwedge_{e=(u,v)\in E^P}
    \mathsf{Adm}_t(e,a) \\
    &=
    \bigwedge_{e=(u,v)\in E^P}
    \mathsf{CI}
    \left(
        \left\langle
            d,\,
            u,\,
            v,\,
            a,\,
            \rho^P(e)
        \right\rangle
        \,\middle|\,
        D_t(v)
    \right).
\end{aligned}
\label{eq:tgci-expanded}
\end{equation}
Thus, $\mathsf{TGCI}(\Phi,\mathcal{S}_t)=1$ only if the entire
rooted propagation scope is admissible. Although the underlying CI
predicate is evaluated at each atomic boundary, the TG-CI judgment is
graph-level because the propagation scope $P$ is part of the proposed
flow $\Phi$ and the flow is accepted only when all transfers in $P$
are appropriate.

\section{Operationalization}
\paragraph{Architecture overview.}
Our implementation treats TG-CI as a binary judge and separates it from the mechanisms used to construct and enforce a propagation region. For each memory item $m_i$, the disclosure controller returns either $(\textsc{Release}, P_i)$, where $P_i$ is the authorized propagation scope, or $(\textsc{Block}, \varnothing)$. We instantiate this controller in two modes, as shown in Figure \ref{fig:workflow}. Full-Scope Mode evaluates the item over the complete downstream MATG, whereas Auto-Region Mode constructs the maximal rooted region through which the item remains admissible. Both modes use the same TG-CI formalism and differ only in how $P_i$ is selected. We assume that the MATG remains fixed throughout each task execution and leave supporting dynamic replanning to future work.
%We assume that the MATG remains fixed throughout each task execution; supporting dynamic replanning would require graph-version tracking and reauthorization of affected propagation scopes, which we leave to future work.

\paragraph{Item-level authorization and exposure-state tracking.}
A memory response is first decomposed into semantic items
$\mathcal{R}(q_{r_0})=\{m_1,\ldots,m_n\}$, following the memory
representation in Section~\ref{sec:ps}. For the disclosure state $D_t$, the implementation maintains a conservative
authorization state $\overline{D}_t$ (Appendix \ref{app:conservative-disclosure}), where
$\overline{D}_t(v)$ includes information actually delivered to $v$ and
items currently authorized to reach $v$. Each memory response is
evaluated using a transaction-local state $\widehat{D}$ initialized
from $\overline{D}_t$. Whenever $m_i$ is tentatively authorized over $P_i$, it is added to
$\widehat{D}(v)$ for every $v\in V(P_i)$ before the next item is
evaluated. This captures cumulative disclosure
both within the current response and across previously authorized
flows. If the response is authorized, the tentative updates are
committed to $\overline{D}_t$; otherwise, they are discarded.

\paragraph{Full-scope mode.}
For an item $m_i$, let $P_i^{\mathrm{full}}$ denote the complete
downstream closure of the requesting agent $r_0$, augmented with the
initial memory-release edge $(s,r_0)$. The controller returns
$(\textsc{Release},P_i^{\mathrm{full}})$ only if
\begin{equation}
    \mathsf{TGCI}
    \left(
        \langle d_i,a_i,P_i^{\mathrm{full}}\rangle,
        \widehat{\mathcal S}_t
    \right)
    =1;
\end{equation}
otherwise, it returns $(\textsc{Block},\varnothing)$. Consequently,
any inadmissible downstream transfer blocks the item before it enters
the workflow. This provides the strongest anticipatory containment
among our two modes, but is deliberately conservative: a distant
recipient can prevent release even when the immediate recipient could
appropriately use the item.

\paragraph{Auto-region mode.}
Auto-Region Mode avoids allowing a distant incompatibility to block an
otherwise appropriate current disclosure. For each item, it constructs
the maximal rooted propagation region
$P_i^\star\subseteq P_i^{\mathrm{full}}$ reachable from the memory
source through contextually admissible edges. If the initial
memory-to-agent edge is inadmissible, the item is blocked; otherwise,
the controller returns $(\textsc{Release},P_i^\star)$. Expansion stops
at the first inadmissible edge on each downstream path, producing the
item-specific propagation frontier
\begin{equation}
    \partial P_i^\star
    =
    \left\{
        (u,v)\in E(P_i^{\mathrm{full}})
        \;\middle|\;
        u\in V(P_i^\star),
        \ (u,v)\notin E(P_i^\star)
    \right\}.
\end{equation}
By construction, $P_i^\star$ satisfies TG-CI, while workflow stages
outside the region may continue after the protected item has been
removed.

\paragraph{Frontier-aware communication gate.}
Each active authorization installs an item-specific frontier rule in a
lightweight communication gate. For sender $u$, the gate consults its
outgoing neighborhood in the current MATG, $\mathcal{N}_{G}^{+}(u)=\{v\in V \mid (u,v)\in E\}$,
and maintains a filter map $\mathcal{F}_u(v)$ containing protected
items prohibited on edge $u\to v$. Messages to recipients
$v\notin\mathcal{N}_{G}^{+}(u)$ are blocked. An on-graph message passes
without semantic inspection when $\mathcal{F}_u(v)=\varnothing$;
otherwise, the gate applies semantic claim detection only for the
listed items and removes matching content, blocking the message if
safe separation is not possible. Thus, expensive content inspection
is concentrated at TG-CI frontiers rather than repeated at every
communication boundary.

\section{Experiments}
\subsection{MAPLE-Bench}
\paragraph{Motivation for a complementary benchmark.}
Recent benchmarks have expanded privacy evaluation beyond final outputs,
but emphasize different dimensions of multi-agent leakage.
AgentSCOPE localizes contextual violations across intermediate
agent--tool flows \citep{arXiv2026Ngong}, while AgentLeak audits
leakage through internal channels, including inter-agent messages and
shared memory \citep{access2026Yagoubi}. PiSAs annotates both task
appropriateness and recipient visibility across agent topologies and
memory regimes \citep{bench_pisas}; MAMA isolates the effect of network
topology on adversarial memory extraction \citep{bench_mama}; and
\emph{The Sum Leaks More Than Its Parts} studies sensitive inferences
formed by composing individually benign disclosures
\citep{arxiv2025patil}. These benchmarks establish important individual
risk dimensions, but do not provide a single
controlled setting that jointly varies MATG topology, recipient
disclosure history, compromised-agent placement, and item-level
propagation permissions while retaining an objective measure of task
completion. MAPLE-Bench
complements this literature by combining these factors within the same
cases, enabling controlled evaluation of whether a defense can preserve
task utility while preventing direct, cumulative, and transitive
disclosure.

%We introduce \textbf{MAPLE-Bench} (Multi-Agent Privacy Leakage Evaluation),
%a controlled benchmark of synthetic multi-agent workflows with
%recipient-specific permissions, annotated forbidden information
%%%%%combinations, and communication-graph variants.
%%%%Tasks, facts, permissions, and answer keys remain fixed across topology
%%%variants, with validation of authorized task-completion routes.
%%Workflows include private and shared memory and compromised-agent
%configurations.
%Fixed reference annotations, hidden from evaluated defenses, support
%separate measurement of recipient-level unauthorized disclosure,
%cumulative exposure, and task success.
%This protocol supports controlled comparisons of privacy defenses
%on the same underlying tasks, including their ability to contain
%downstream propagation without unnecessarily restricting useful
%information flow.

\paragraph{Benchmark composition.}
MAPLE-Bench contains 144 cases across four application scenarios, with 36 cases per scenario. Each case instantiates a four-agent MATG and a terminal node representing the user or final-output recipient. Within each scenario, cases are evenly divided among chain, tree, and branching topologies. 108 cases contain one compromised agent and 36 contain two, with compromised-agent locations varied across the MATG. Each case contains 30 synthetic memory items with fictitious values: 12 agent-specific private items, with three associated with each task agent, and 18 shared items. Only seven items contribute to the answers of five multiple-choice questions, either individually or in combination; the remaining 23 items serve as distractors to model memory stores in which most recorded information is irrelevant to the current task, simulating the real-life scenario. The detail of the benchmark is provided in Appendix~\ref{app:benchmark}.

\paragraph{Privacy annotations and evaluation.}
Each case provides a task-conditioned item--recipient policy matrix
$Y$, where $Y_{i,v}=1$ iff recipient $v$ may receive memory item $m_i$
for its assigned task. Policies cover all four agents and the terminal
output recipient, providing ground truth for direct disclosures and
for transitive violations. Cases targeting cumulative leakage additionally
identify item combinations whose joint disclosure reveals a protected
fact. Each benchmark instance therefore specifies the MATG, agent
subtasks, compromised-agent configuration, memory items, privacy
policies, multiple-choice questions, and gold answers. Because all
sensitive values are synthetic, MAPLE-Bench enables
measurement of task accuracy, unauthorized and cumulative exposure, and propagation
containment without using real personal data.

\subsection{Baselines and Experimental Conditions}
\paragraph{Baselines.}
We compare the two TG-CI enforcement modes against three baselines. \textbf{No Protection} returns all retrieved memory items to the
requesting agent and delivers inter-agent messages without privacy
judgment or filtering. \textbf{Local-History CI} retains item-level
judgment and recipient disclosure-state tracking, and evaluates each
item over the initial memory-release edge. 
%Specifically, for
%requester $r_0$, it 
%uses the single-edge propagation scope and 
%releases $m_i$ only if
%$\mathsf{CI}
%    \left(
%        \left\langle
%        d,\,
%        \mathcal{M},\,
%        r_0,\,
%        a,\,
%        \rho(e_{(\mathcal{M},r_0)})
%    \right\rangle
%        \mid
%        D_t(r_0)
%    \right)=1$.
%This baseline captures direct and cumulative
%disclosure risk at the current recipient, but neither considers
%downstream MATG structure nor installs frontier filters. 
Public
implementations of the closest cumulation-aware systems, OCELOT and
CAMP, were unavailable at the time of evaluation
\citep{arxiv2026xie,arxiv2026Panjwani}; we therefore use Local-History
CI as a controlled in-framework comparison rather than claiming an
exact reproduction. We also evaluate \textbf{Online Atomic CI}, which  applies the same edge-level atomic CI used by TG-CI independently at every realized memory and inter-agent transfer.  The proposed systems are evaluated as
\textbf{TG-CI Full-Scope} and \textbf{TG-CI Auto-Region}.

\paragraph{Experimental conditions.}
We implement the multi-agent workflows using LangChain and use Mem0 as
the persistent memory backend. Task agents perform memory retrieval,
inter-agent communication, and final-answer generation, while a
separate privacy judge evaluates item- and edge-level contextual
admissibility. We evaluate three agent--judge model assignments:
\textsc{Luna--Luna}, \textsc{Haiku--Haiku}, and
\textsc{Haiku--Luna}, where the two names denote the task-agent and
privacy-judge models, respectively. GPT-5.6 Luna and Claude Haiku 4.5 are
used as the underlying model families. Unless otherwise specified,
all main-text results are reported for the \textsc{Haiku--Luna}
assignment, with results for remaining agent--judge configurations provided in Appendix ~\ref{app:additional-result}. 
%Exact API identifiers and
%decoding settings are reported in Appendix~\ref{app}.
Frontier filtering is performed locally using MiniLM for semantic
candidate retrieval and DeBERTa for bidirectional entailment
verification. 
%These models are invoked only when an inter-agent edge
%has a non-empty filter list; other on-graph messages follow the gate's
%fast path without semantic inspection.
Details of this procedure are
provided in Appendix~\ref{sem_check}.

\paragraph{Metrics.}
We report four run-level metrics. \textbf{Unauthorized Disclosure
Count} is the number of delivery events in which a memory item reaches
a recipient for whom it is inappropriate according to the benchmark
policy; repeated disclosures are counted separately. For cumulative
cases, the delivery that first completes an annotated prohibited item
combination is counted as an unauthorized disclosure.
\textbf{Compromised-Agent Exposure Count} is the number of distinct
inappropriate item observed by compromised agents,
counting repeated delivery of the same item to the same agent only
once. \textbf{Judge Execution Token Count} is the total number of additional input
and output tokens used by  privacy-judge LLM calls;
local MiniLM and DeBERTa inference is excluded. \textbf{Utility Score}
is the fraction of the five multiple-choice questions answered
correctly.

\subsection{Privacy--Cost Trade-off}
\begin{figure}[htbp]
    \centering
    % Top-Left Subfigure
    \begin{subfigure}[b]{0.49\linewidth}
        \centering
        \includegraphics[trim=5 6 5 5, clip,width=\textwidth]{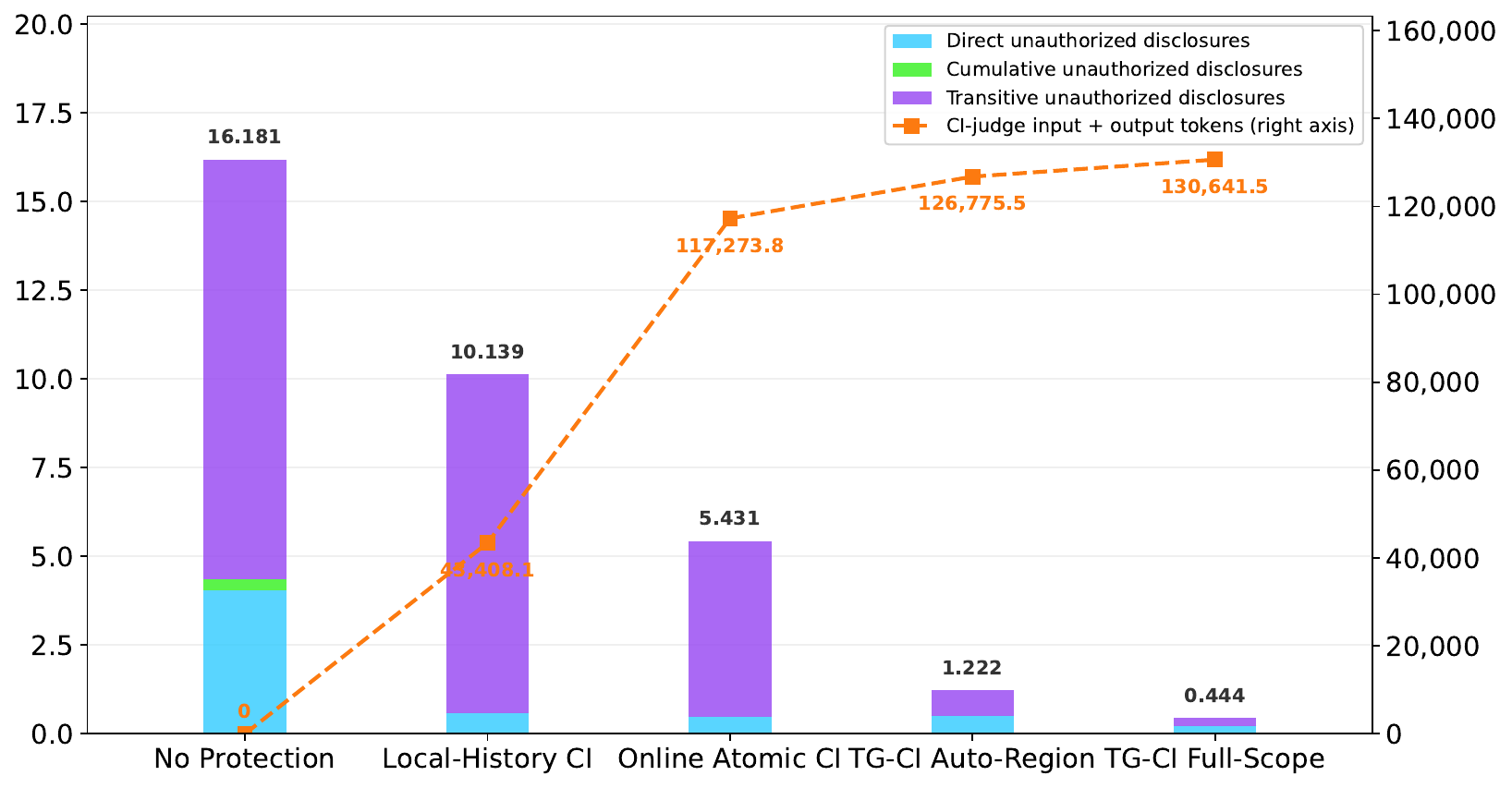}
        \caption{Healthcare}
        \label{fig:sub_a}
    \end{subfigure}
    \hfill
    % Top-Right Subfigure
    \begin{subfigure}[b]{0.49\linewidth}
        \centering
        \includegraphics[trim=5 6 5 5, clip,width=\textwidth]{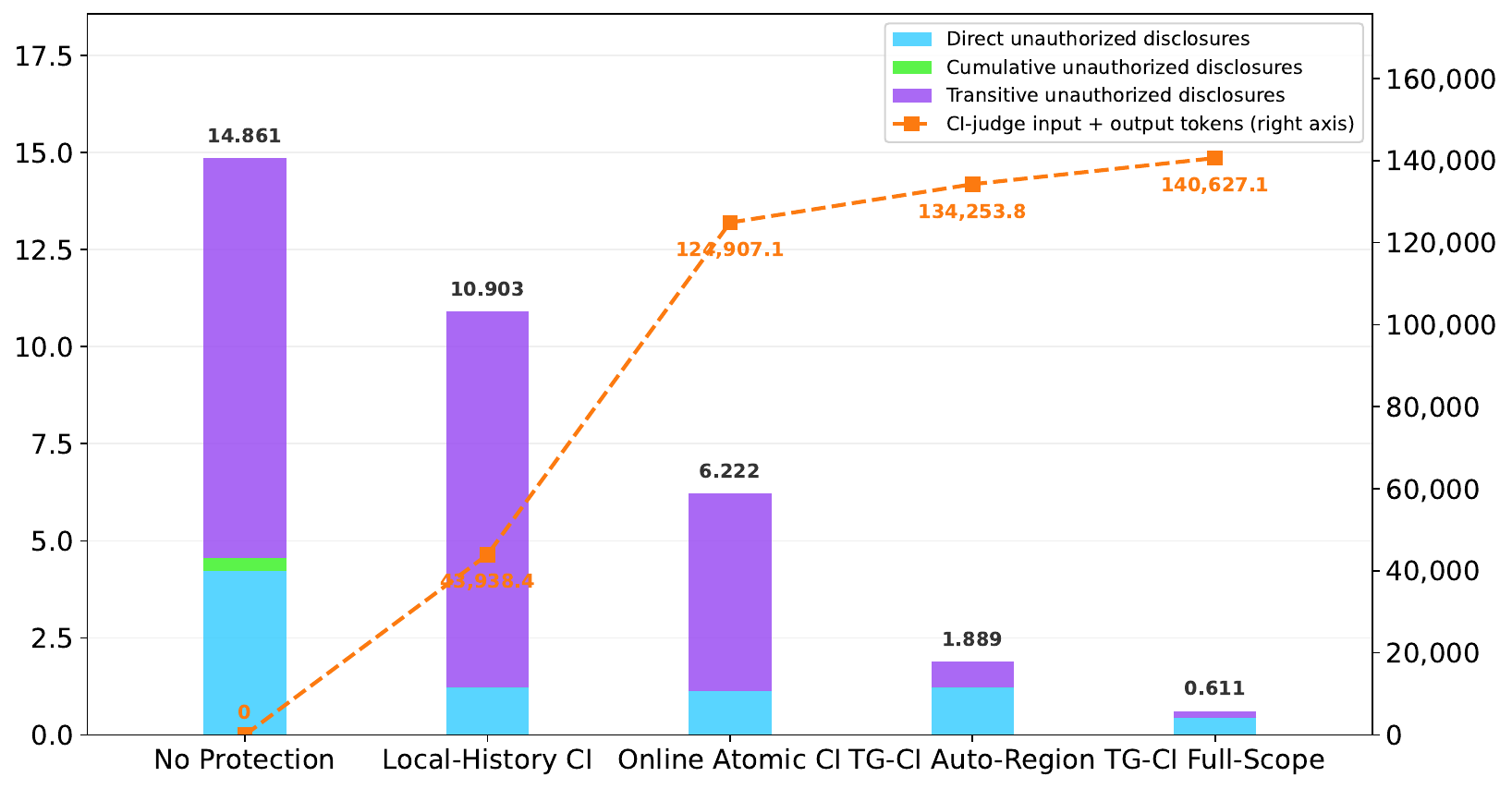}
        \caption{Finance}
        \label{fig:sub_b}
    \end{subfigure}

    \vspace{0.2em} % Adjust vertical spacing between rows as needed

    % Bottom-Left Subfigure
    \begin{subfigure}[b]{0.49\linewidth}
        \centering
        \includegraphics[trim=5 6 5 5, clip,width=\textwidth]{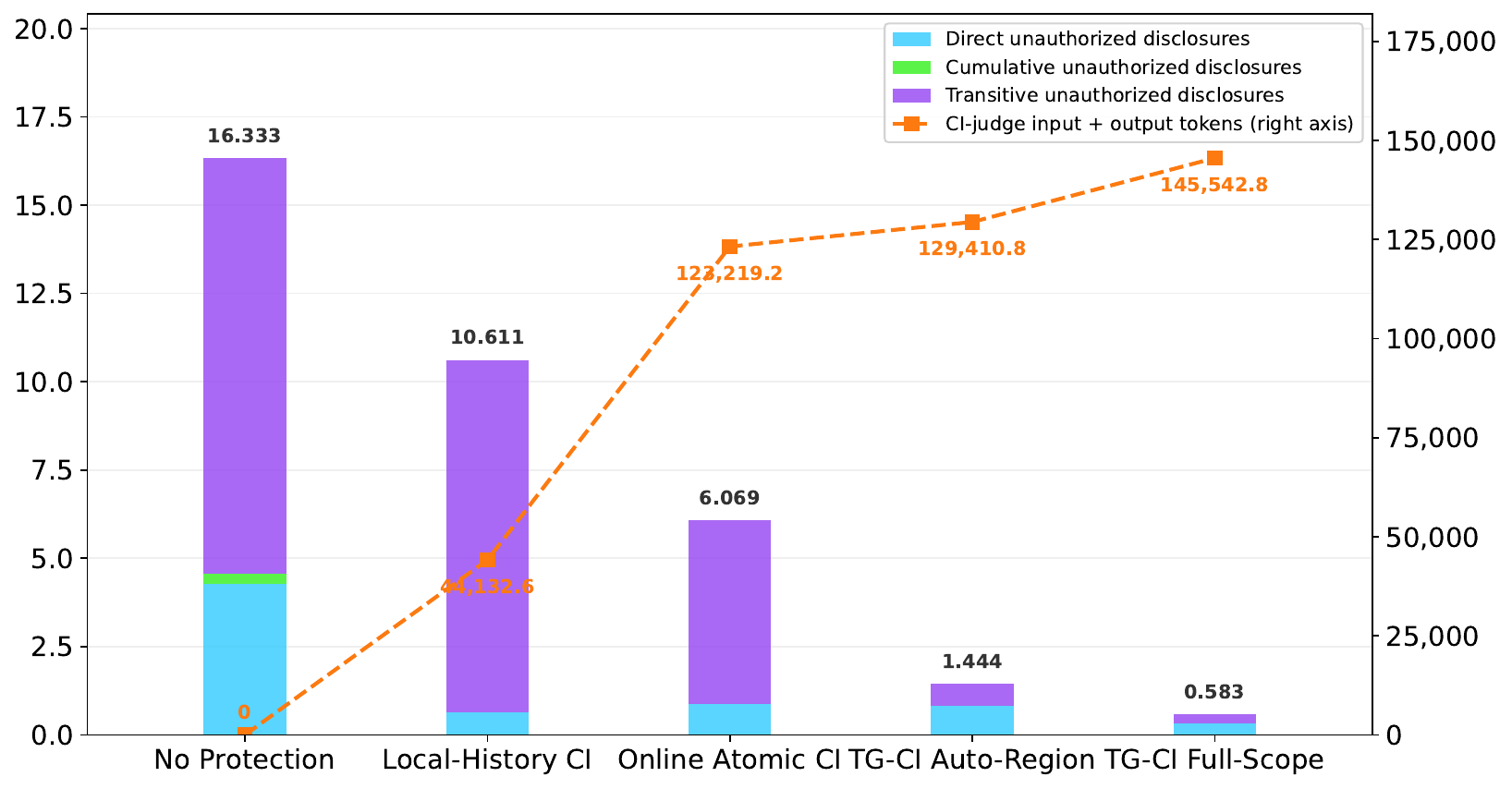}
        \caption{HR Affairs}
        \label{fig:sub_c}
    \end{subfigure}
    \hfill
    % Bottom-Right Subfigure
    \begin{subfigure}[b]{0.49\linewidth}
        \centering
        \includegraphics[trim=5 6 5 5, clip,width=\textwidth]{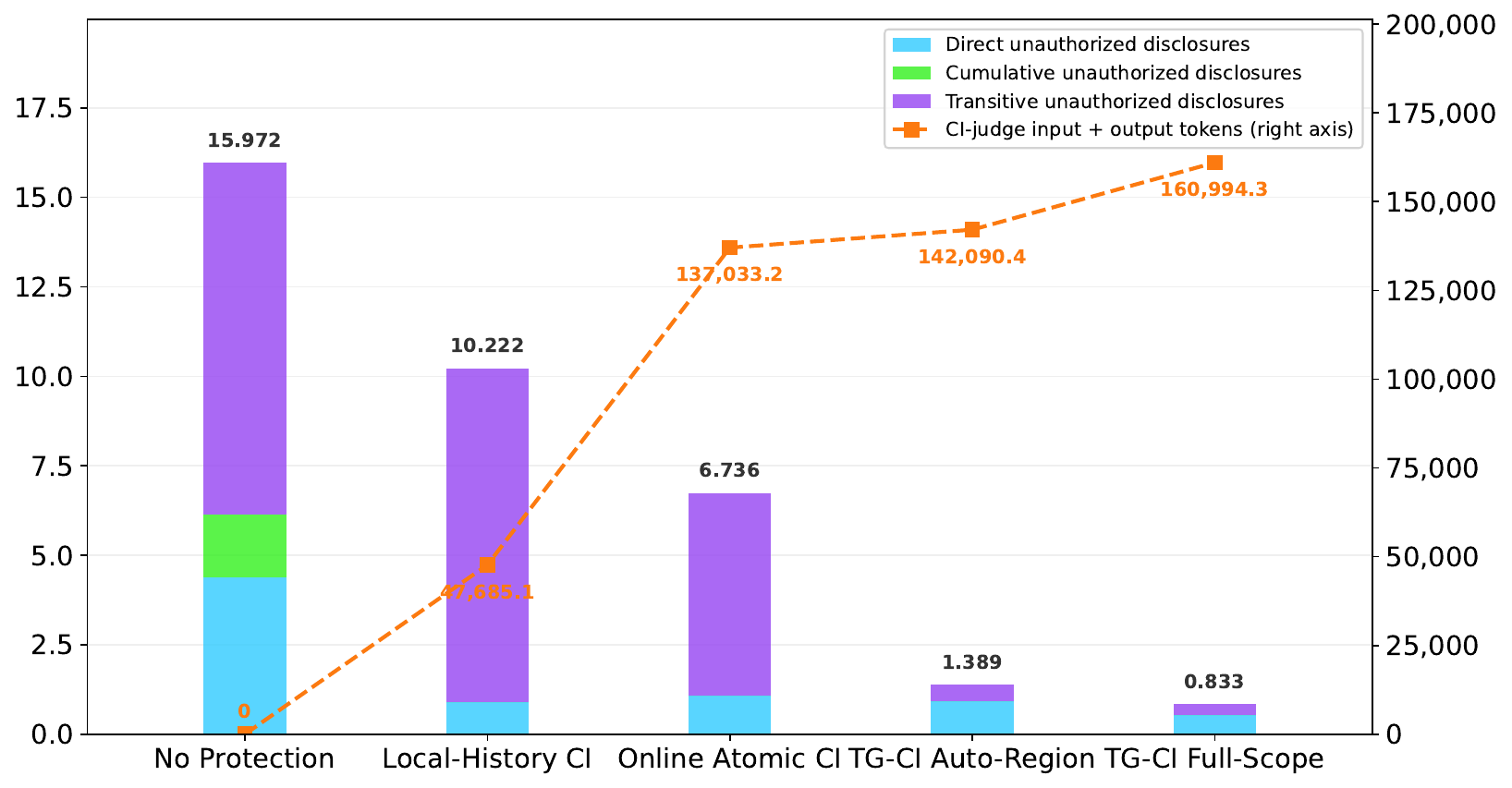}
        \caption{Lifestyle}
        \label{fig:sub_d}
    \end{subfigure}

    \caption{Privacy--cost trade-off across four application scenarios. Stacked bars decompose the mean Unauthorized Disclosure Count into mutually exclusive direct, cumulative, and transitive violations (left axis). The dashed line reports Judge Execution Token Count (right axis).}
    \label{fig:privacy_cost_tradeoff}
\end{figure}
Figure~\ref{fig:privacy_cost_tradeoff} shows a consistent privacy--cost trade-off across all
four scenarios: \textbf{both TG-CI variants incur higher computation than the baselines, but achieve markedly stronger privacy protection.} Averaged across scenarios, the unauthorized disclosure
count decreases from 15.84 under No Protection to 10.47 under
Local-History CI and 6.11 under Online Atomic CI. TG-CI further reduces the count to 1.49 under Auto-Region and
0.62 under Full-Scope. This improvement comes with increased LLM usage: Mean CI-judge token consumption rises to 133.1~K for Auto-Region Mode and
144.5~K for Full-Scope Mode. 
%In particular, Full-Scope reduces
%unauthorized disclosure by 94.1\% relative to Local-History CI while
%using approximately 15.1\% more judge tokens.

The disclosure-type decomposition clarifies where these gains arise. All three violation types occur under No Protection, with transitive disclosures forming the largest component. Local-History CI reduces direct and cumulative violations but leaves substantial transitive leakage. Online Atomic CI further reduces each category, yet transitive disclosure remains its dominant residual source. Both TG-CI variants substantially reduce direct, cumulative, and transitive violations, with their largest advantage over the atomic baselines appearing in the transitive component. These results suggest that atomically checking realized transfers is insufficient in practice and \textbf{prospectively reasoning over the downstream MATG with either whole-scope admission or maximal admissible-region authorization enforcement provides additional protection against
transitive leakage.} Categories are mutually exclusive with cumulative classification takes precedence whenever a delivery completes an annotated prohibited combination.

Auto-Region and Full-Scope occupy different points on this frontier. Auto-Region retains a small residual disclosure count while requiring approximately 7.8\% fewer judge tokens than Full-Scope. Full-Scope achieves the lowest disclosure count by requiring the complete downstream closure to be admissible. The reported Full-Scope %values are the measured execution costs rather than an upper-bound estimate. Its 
slightly higher cost reflects stricter source-level blocking, which induces additional memory requests by downstream task agents in workflows.

\subsection{Compromised-Agent Exposure}
\begin{figure}[htbp]
    \centering
    \includegraphics[width=\linewidth]{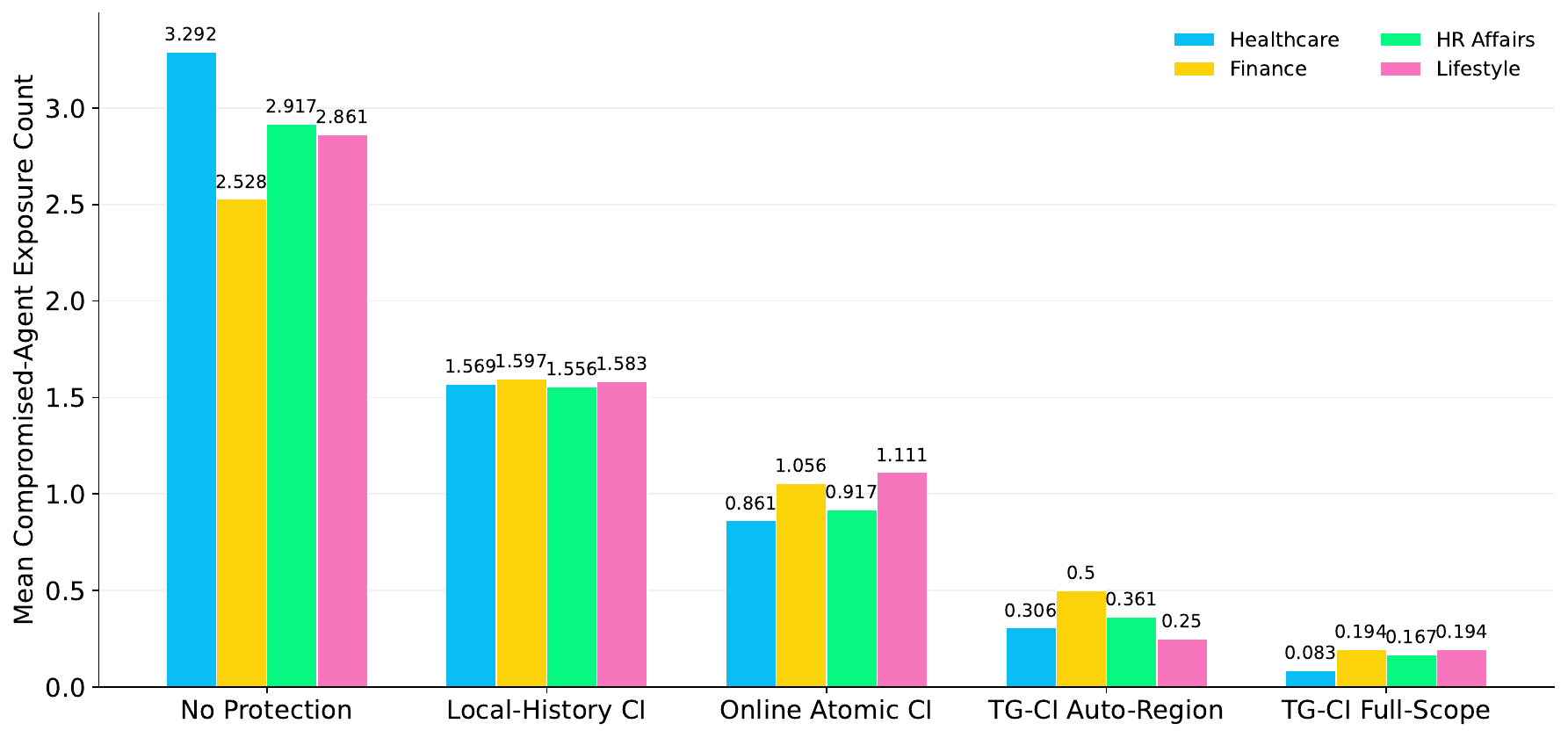}
    \vspace{-5mm}
    \caption{Mean number of distinct inappropriate memory items exposed to compromised agents across four application scenarios. 
    %Both TG-CI modes consistently reduce adversarial exposure relative to No Protection and Local-History CI.
    }
    \label{fig:compromised-agent-exposure}
    \vspace{-2mm}
\end{figure}

Figure~\ref{fig:compromised-agent-exposure} evaluates whether the reduction in disclosure events also limits the distinct protected information acquired by compromised agents. Averaged across the four scenarios, the mean exposure count decreases from 2.90 under No Protection to 1.58 under Local-History CI and 0.99 under Online Atomic CI. TG-CI further reduces exposure to 0.35 under Auto-Region and 0.16 under Full-Scope.
The metric counts distinct exposed items rather than repeated disclosure events. Therefore, this result shows that TG-CI does not merely reduce repeated violation events: it \textbf{effectively limits the distinct protected information available to the compromised coalition}. Full-Scope provides the strongest exposure reduction, while Auto-Region retains a higher exposure in exchange for its lower computational cost.

\subsection{Utility Evaluation}
\begin{table}[htbp]
\centering
\small
\setlength{\tabcolsep}{3.5pt} % Reduce column padding to fit within margins
\resizebox{\linewidth}{!}{% Ensures table stays strictly within page margins
\begin{tabular}{l clclcl}
\toprule
 & \multicolumn{2}{c}{\textbf{\textsc{Haiku--Luna}}} & \multicolumn{2}{c}{\textbf{\textsc{Haiku--Haiku}}} & \multicolumn{2}{c}{\textbf{\textsc{Luna--Luna}}} \\
\cmidrule(lr){2-3} \cmidrule(lr){4-5} \cmidrule(lr){6-7}
\textbf{Condition} & \textbf{Utility Score} & \textbf{vs. NP / LH-CI} & \textbf{Utility Score} & \textbf{vs. NP / LH-CI} & \textbf{Utility Score} & \textbf{vs. NP / LH-CI} \\
\midrule
No Protection     & 80.97\% & (---,97.25\%)    & 81.04\% & (---,103.82\%)    & 94.72\% & (---,98.55\%)    \\
Local-History CI  & 83.26\% & (102.83\%,---) & 78.06\% & (96.32\%,---) & 96.11\% & (101.47\%,---) \\
Online Atomic CI  & 80.69\% & (99.65\%,96.91\%) &74.04\% & (91.36\%,94.85\%) &88.47\% &(93.40\%,92.05\%)\\
TG-CI Auto-Region & 80.00\% & (98.80\%,96.08\%) & 68.40\% & (84.40\%,87.62\%) & 92.22\% & (97.36\%,95.95\%) \\
TG-CI Full-Scope  & 77.92\% & (96.23\%,93.58\%) & 62.22\% & (76.78\%,79.72\%) & 88.06\% & (92.96\%,91.62\%) \\
\bottomrule
\end{tabular}%
}
\caption{Task utility across agent--judge model assignments. Utility Score denotes mean multiple-choice accuracy in percentage points. Values in parentheses report score relative to No Protection and Local-History CI, respectively.}
\label{tab:utility_evaluation}
\vspace{-4mm}
\end{table}

Table~\ref{tab:utility_evaluation} shows that TG-CI's privacy gains do
not require indiscriminately removing task-relevant information.
Across the three agent--judge assignments, Auto-Region retains
84.4--98.8\% of No Protection utility and 87.6--96.1\% of
Local-History CI utility. Full-Scope retains 76.8--96.2\% and
79.7--93.6\%, respectively. Together with the substantial leakage
reductions reported above, these results indicate that
\textbf{TG-CI primarily suppresses contextually inappropriate
disclosure while retaining information needed for task completion}. 

The two TG-CI modes also provide a clear privacy--utility trade-off. Auto-Region consistently outperforms Full-Scope by 2.1--6.2 percentage. \textbf{Full-Scope therefore provides stronger containment at the cost of potentially more aggressive removal of potentially useful information}. Online Atomic CI generally achieves a similar utility performance as TG-CI Auto-Region and fluctuates depending on specific assignments.

\subsection{MATG Topology}
\begin{figure}[htbp]
    \centering
    % Subfigure A
    \begin{subfigure}[b]{0.24\linewidth}
        \centering
        \includegraphics[trim=7 10 5 18, clip,width=\textwidth]{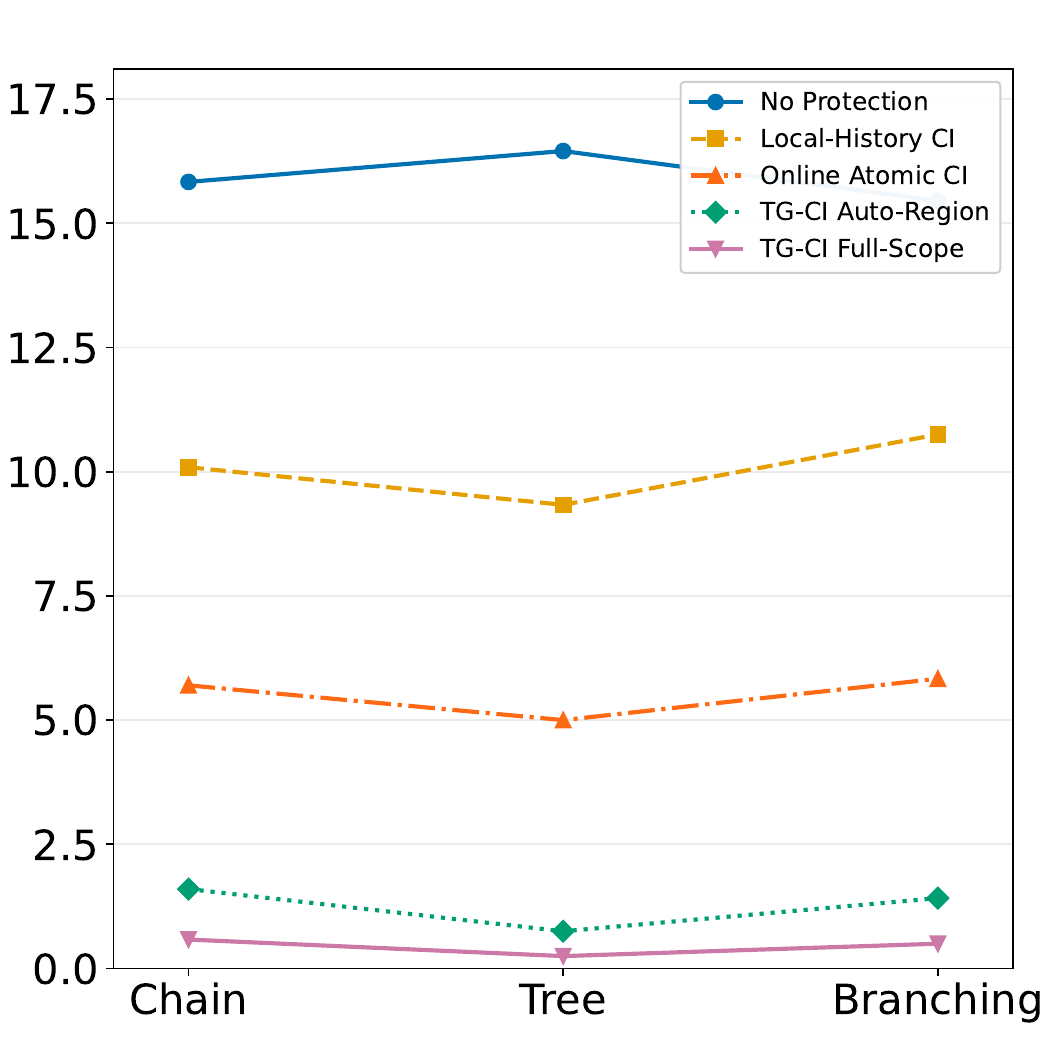}
        \caption{Healthcare}
        \label{fig:sub_a}
    \end{subfigure}%
    \hfill
    % Subfigure B
    \begin{subfigure}[b]{0.24\linewidth}
        \centering
        \includegraphics[trim=2 8 2 16, clip,width=\textwidth]{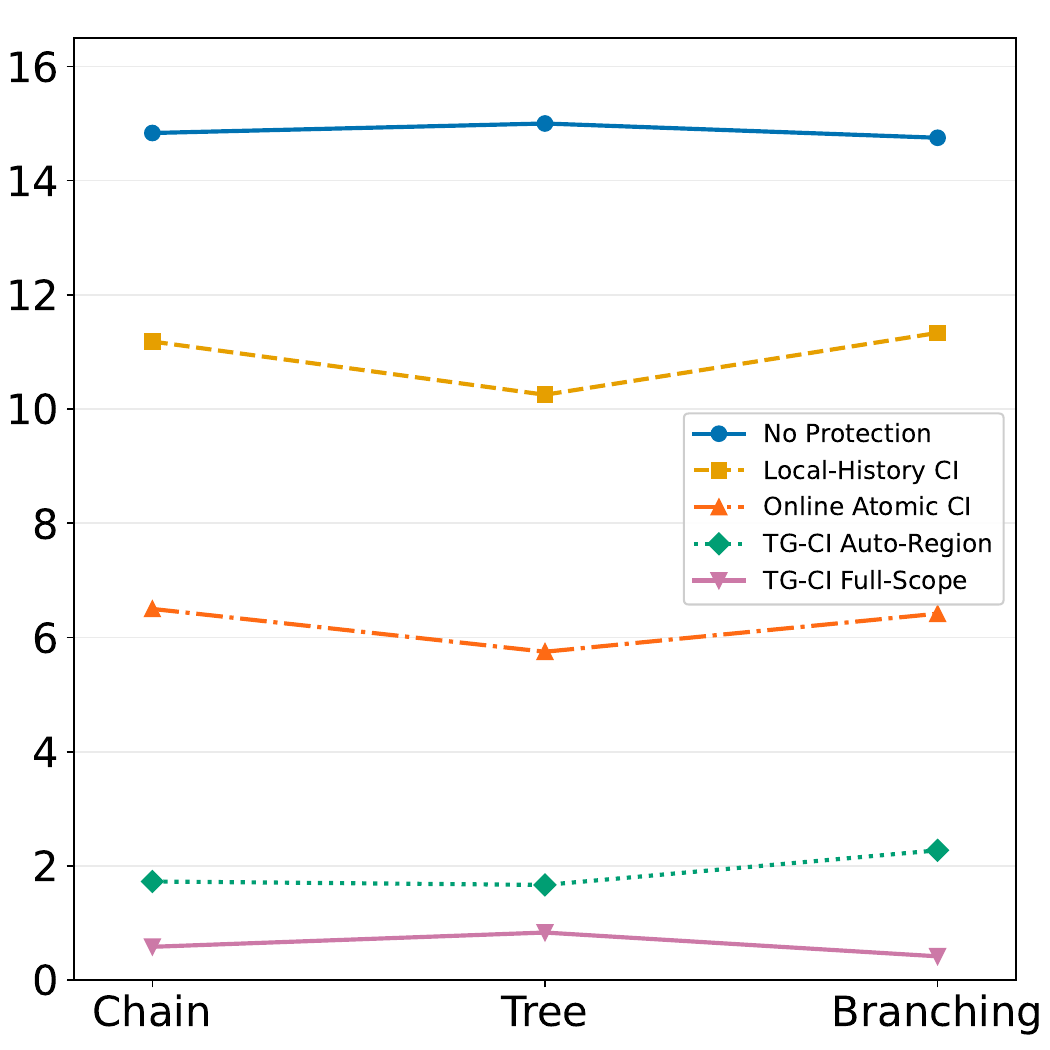}
        \caption{Finance}
        \label{fig:sub_b}
    \end{subfigure}%
    \hfill
    % Subfigure C
    \begin{subfigure}[b]{0.24\linewidth}
        \centering
        \includegraphics[trim=7 10 5 18, clip,width=\textwidth]{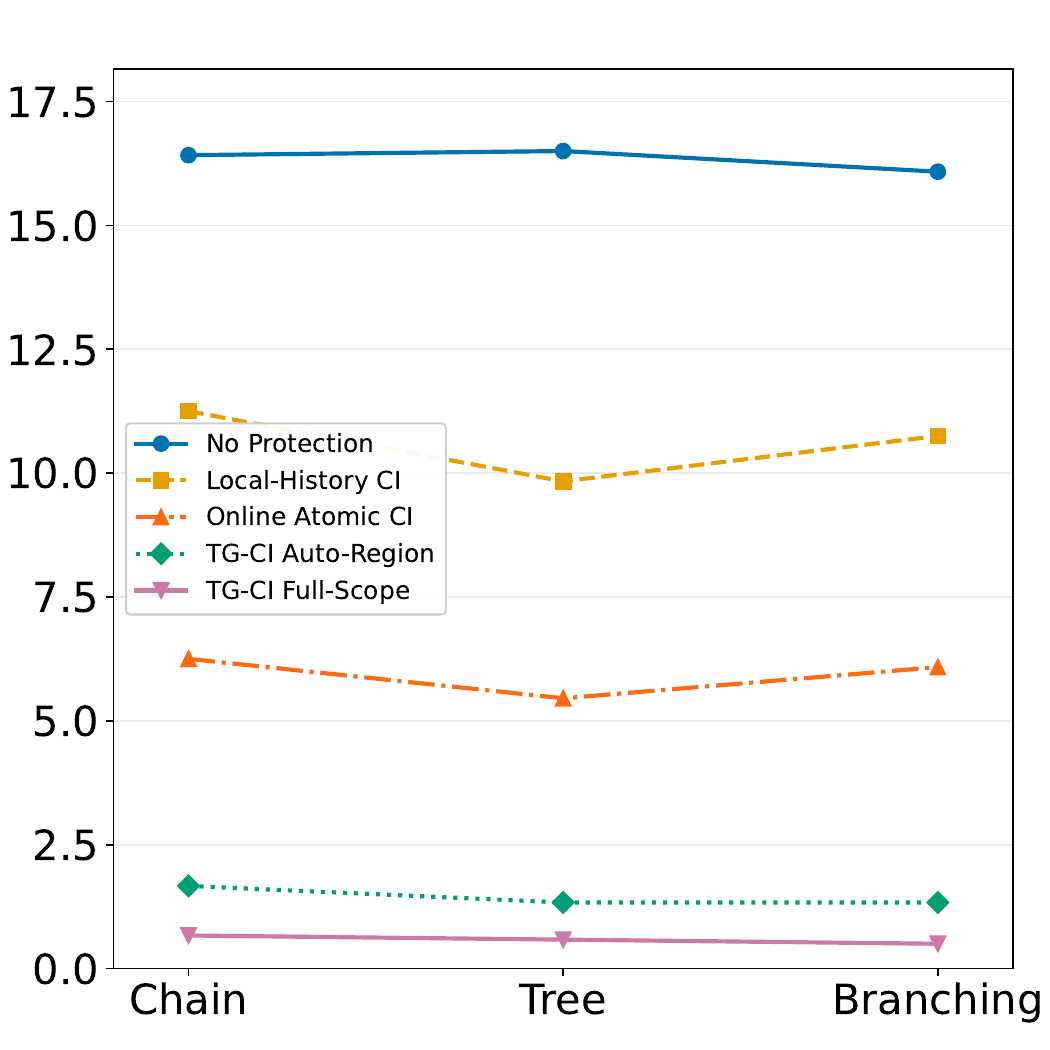}
        \caption{HR Affairs}
        \label{fig:sub_c}
    \end{subfigure}%
    \hfill
    % Subfigure D
    \begin{subfigure}[b]{0.24\linewidth}
        \centering
        \includegraphics[trim=5 10 5 18, clip,width=\textwidth]{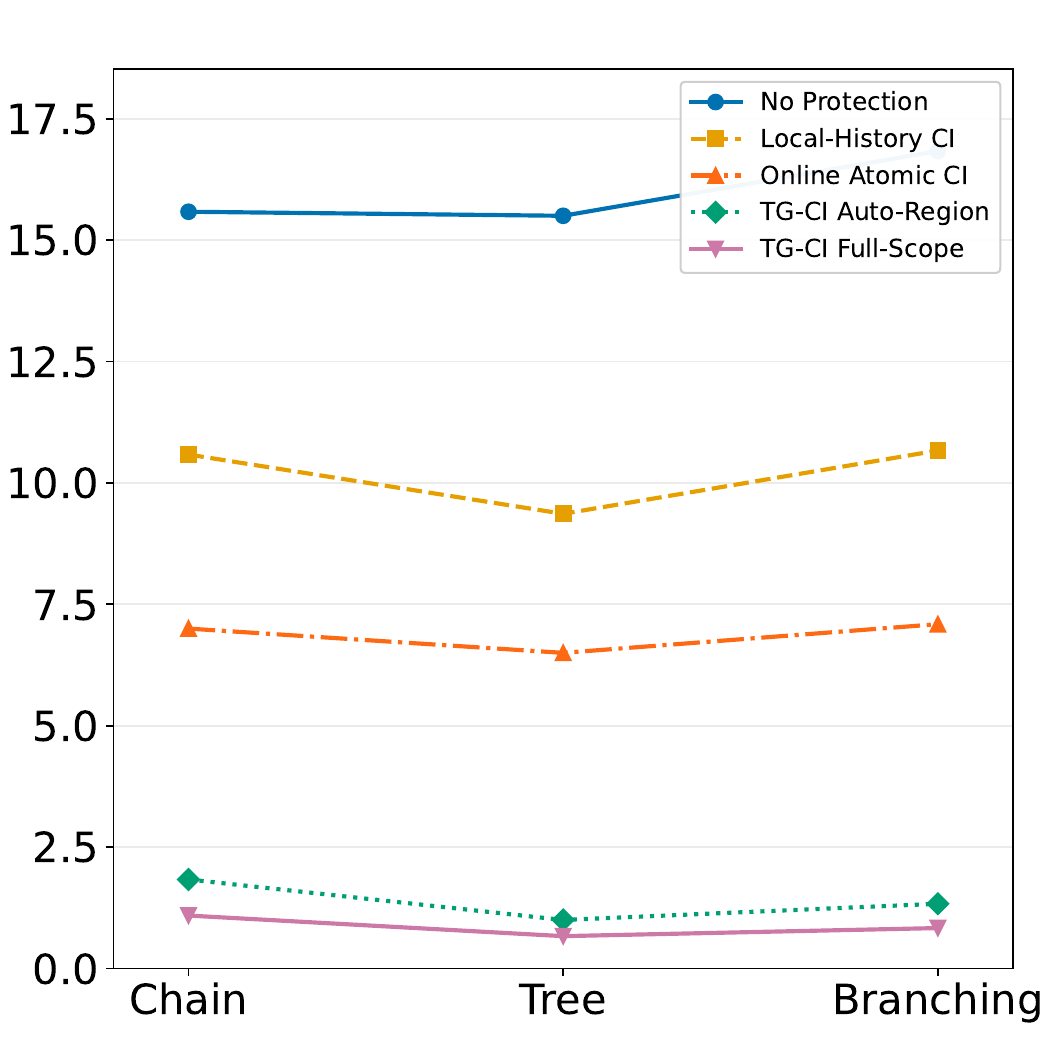}
        \caption{Lifestyle}
        \label{fig:sub_d}
    \end{subfigure}

    \caption{Mean unauthorized disclosure count (left axis) across chain, tree, and branching MATG topologies for four application scenarios.}
    \label{fig:MATG_Topology}
    \vspace{-2mm}
\end{figure}

Figure~\ref{fig:MATG_Topology} examines unauthorized disclosure across
chain, tree, and branching MATGs. The relative ordering of the methods
is consistent across all four application scenarios: No Protection
produces the most violations, Local-History CI and Online Atomic CI provides a moderate
reduction, and both TG-CI variants maintain substantially lower
disclosure counts. In particular, TG-CI Auto-Region and Full-Scope show
only limited variation across the three topologies, with Full-Scope
remaining the strongest protection in every setting. Although
branching structures introduce additional potential propagation paths,
we observe no systematic degradation in TG-CI performance. These
results suggest that \textbf{the privacy improvements of TG-CI are robust
across the evaluated MATG structures rather than being specific to a
particular topology}.

\section{Conclusion}
We introduced \textbf{Task-Graph Contextual Integrity (TG-CI)}, a task-graph-native extension of Contextual Integrity that evaluates protected disclosures as stateful information flows over a Multi-Agent Task Graph, jointly accounting for recipient disclosure histories and intended downstream propagation. We operationalized TG-CI through propagation-scoped authorization with two complementary policies: Full-Scope Mode for conservative end-to-end protection and Auto-Region Mode for lower-cost frontier-based containment. We further introduced MAPLE-Bench, a controlled benchmark for direct, cumulative, and transitive privacy violations in multi-agent workflows. 
Our experiments show TG-CI effectively reduced unauthorized disclosure and compromised-agent exposure while retaining substantial task utility. 
These results demonstrate that moving beyond local disclosure checks toward graph-aware, stateful privacy reasoning can substantially contain information leakage while preserving useful multi-agent collaboration.

\subsection*{AI use statement}
In this work, we used generative AI tools to generate synthetic data sets. We have not used generative AI tools to help develop theoretical models or conceptual frameworks, formulate mathematical claims, provide critical ingredients for proving mathematical claims, assist in the writing of proofs, propose or refine hypotheses, design or provide feedback on research methodology or experiments, implement methods, assist with translation, clean and reformat datasets, support qualitative and thematic data analysis, or interpret results. We have reviewed all AI-assisted work. All the synthetic dataset and notation is human-reviewed by the researcher. We take responsibility for the final content of this work, including text, claims, or artifacts produced with the aid of generative AI.

\subsection*{Ethics statement}
This work studies privacy risks and defenses in multi-agent systems but does not involve human participants or real personal records. All identities, memory items, sensitive attributes, and task scenarios in MAPLE-Bench are synthetically generated and use fictitious values that are not intended to correspond to actual individuals. The benchmark and data also include task-specific privacy policies, which may encode incomplete or culturally contingent assumptions about appropriate information flow. We consequently do not claim that these policies represent universal privacy norms, and deployment in real settings would require domain expertise, stakeholder input, and compliance with applicable legal and organizational requirements.

\subsection*{Reproducibility statement}
To support reproducibility, we provide our source code and MAPLE-Bench as supplementary materials accompanying this submission. Sections~2--4 specify the problem setting, TG-CI formalization, and the Full-Scope and Auto-Region enforcement procedures. Section~5.1 describes the benchmark composition and privacy annotations, while Section~5.2 presents the baselines, experimental conditions, and evaluation metrics.

%\subsubsection*{Acknowledgments}
%Use unnumbered third level headings for the acknowledgments. All
%acknowledgments, including those to funding agencies, go at the end of the paper.

\bibliography{iclr2027_conference}
\bibliographystyle{iclr2027_conference}

\clearpage
\appendix
\section{Conservative Disclosure States}
\label{app:conservative-disclosure}

Let $H_1 \preceq H_2$ mean that $H_1$ contains no more
privacy-relevant information than $H_2$. Since the conservative
authorization state includes both delivered information and items
authorized to reach a recipient, we have
$D_t(v) \preceq \overline{D}_t(v)$.

\paragraph{Assumption (downward-closed admissibility).}
For a fixed atomic flow $\varphi_e(a)$, acceptance under a larger
disclosure footprint implies acceptance under every smaller footprint:
\begin{equation}
H_1 \preceq H_2
\;\land\;
\operatorname{CI}(\varphi_e(a) \mid H_2)=1
\;\Longrightarrow\;
\operatorname{CI}(\varphi_e(a) \mid H_1)=1.
\label{eq:disclosure-downward-closure}
\end{equation}

Under this assumption, evaluating against $\overline{D}_t(v)$,
or its transaction-local extension $\widehat{D}(v)$, is conservative:
acceptance also holds under any actual disclosure state bounded by
the evaluated footprint. Undelivered authorizations may therefore
cause additional rejections, but not false acceptances due solely
to state overapproximation. This justification requires the footprint
used for authorization to remain an upper bound on the recipient's
actual state at delivery.

\section{Benchmark Construction}
\label{app:benchmark}
\begin{table}[H]
\small
\setlength{\tabcolsep}{4pt}
\renewcommand{\arraystretch}{1.05}

% Panel (a): domain-level distribution.
\begin{tabular*}{\linewidth}
    {@{\extracolsep{\fill}}lrrrr@{}}
\toprule
\multicolumn{5}{@{}l@{}}{\textbf{(a) Domain distribution}} \\
\midrule
\textbf{Domain}
& \textbf{Workflows}
& \textbf{Cases}
& \textbf{One attacker}
& \textbf{Two attackers} \\
\midrule
Healthcare   & 12 & 36 & 27 & 9 \\
Finance      & 12 & 36 & 27 & 9 \\
Corporate HR & 12 & 36 & 27 & 9 \\
Lifestyle    & 12 & 36 & 27 & 9 \\
\midrule
\textbf{Total}
& \textbf{48}
& \textbf{144}
& \textbf{108}
& \textbf{36} \\
\bottomrule
\end{tabular*}

\smallskip

% Panel (b): structure, annotations, and reuse.
\begin{tabularx}{\linewidth}{
    @{}
    >{\raggedright\arraybackslash}p{0.29\linewidth}
    >{\raggedright\arraybackslash}X
    @{}
}
\toprule
\multicolumn{2}{@{}l@{}}
    {\textbf{(b) Structure, annotations, and reuse}} \\
\midrule

Graph topologies
& Chain, tree, and branching; 48 cases each. \\

Agents per case
& 4 executing agents and 1 passive answer receiver. \\

Question format
& 5 four-option questions per case:
  3 single-fact and 2 two-fact questions. \\

Question instances
& 720 total; 240 workflow-level questions before
  topology replication. \\

Memory per case
& 30 items: 12 private (3 per executing agent)
  and 18 shared. \\

Memory relevance
& 7 necessary, 7 relevant-only, and 16 unrelated
  items per case. \\

Cumulative-disclosure annotations
& All 144 cases (100\%); 2 sensitive fact-pair rules
  per case; 288 rule instances. \\

Single-attacker positions
& \texttt{P1}: 36 cases;
  \texttt{P2}, \texttt{P3}, and \texttt{P4}: 24 cases each. \\

Two-attacker position pairs
& \texttt{(P1,P2)}, \texttt{(P1,P3)}, and
  \texttt{(P2,P4)}: 12 cases each. \\

Task-template reuse
& Each workflow's task, memory, and complete
  question--answer set are reused across all three topologies. \\

Exact question-prompt reuse
& 237 distinct prompts among the 240 workflow-level questions;
  3 prompts each appear in two workflows. \\

\bottomrule
\end{tabularx}
\centering
\caption{MAPLE-Bench composition and statistics.
Each of the 48 workflows is instantiated under three MATG
topologies, yielding 144 cases. Attacker counts and position
distributions are reported at the case level. \texttt{$P_{i}$} represent agent in \textit{i}-th position.}
\label{tab:maple-bench-statistics}
\end{table}

\subsection{Example Task and Question}
\label{app:two-fact-example}

Table~\ref{tab:maple-two-fact-example} presents a task from
MAPLE-Bench, one associated two-fact question, and its supporting
memory items. Only the two memory items needed for this question
are shown.

\begin{table}[htbp]
\small
\setlength{\tabcolsep}{5pt}
\renewcommand{\arraystretch}{1.08}

\begin{tabularx}{\linewidth}{
    @{}
    >{\raggedright\arraybackslash}p{0.19\linewidth}
    >{\raggedright\arraybackslash}X
    @{}
}
\toprule
\textbf{Component} & \textbf{Content} \\
\midrule

Task
& Please confirm my agreed return date, the first-week hours,
  and whether my fit note is signed. Find a rehabilitation
  review-call slot that fits my availability and the right
  place and day to collect my own paperwork copies. \\

\addlinespace[4pt]

Two-fact question
& Which offered rehabilitation review call slot meets both
  the patient's availability and the service timetable? \\

\addlinespace[4pt]

Answer options
& \textbf{A.} Thursday at 10:30 \newline
  \textbf{B.} Thursday at 15:30 \newline
  \textbf{C.} Tuesday at 15:30 \newline
  \textbf{D.} Tuesday at 10:30 \\

\midrule

Memory item 1
& \texttt{appointment\_availability} \newline
  \emph{Storage:} owner-private;
  \emph{owner:} \texttt{records\_agent}. \newline
  ``For the rehabilitation review call, the patient can attend
  on Thursday only. Either 10:30 or 15:30 is acceptable on
  that day.'' \\

\addlinespace[4pt]

Memory item 2
& \texttt{appointment\_slot\_catalog} \newline
  \emph{Storage:} shared. \newline
  ``The rehabilitation review call service currently offers
  15:30 on both Tuesday and Thursday. It offers no 10:30
  slot on either day.'' \\

\midrule

Correct answer
& \textbf{B: Thursday at 15:30.} \\

\addlinespace[4pt]

Two-fact reasoning
& The availability item leaves options A and B possible,
  whereas the service timetable leaves options B and C
  possible. Their intersection uniquely selects B.
  Neither memory item alone determines the correct option. \\

\bottomrule
\end{tabularx}
\centering
\caption{Example of two-fact question answering in MAPLE-Bench.
The correct answer requires combining the patient's availability
with the service timetable. Memory storage types and ownership
are shown alongside the original item contents.}
\label{tab:maple-two-fact-example}
\end{table}

\section{Additional Results}
\label{app:additional-result}
\begin{figure}[H]
    \centering
    % Top-Left Subfigure
    \begin{subfigure}[b]{0.49\linewidth}
        \centering
        \includegraphics[trim=5 6 5 5, clip,width=\textwidth]{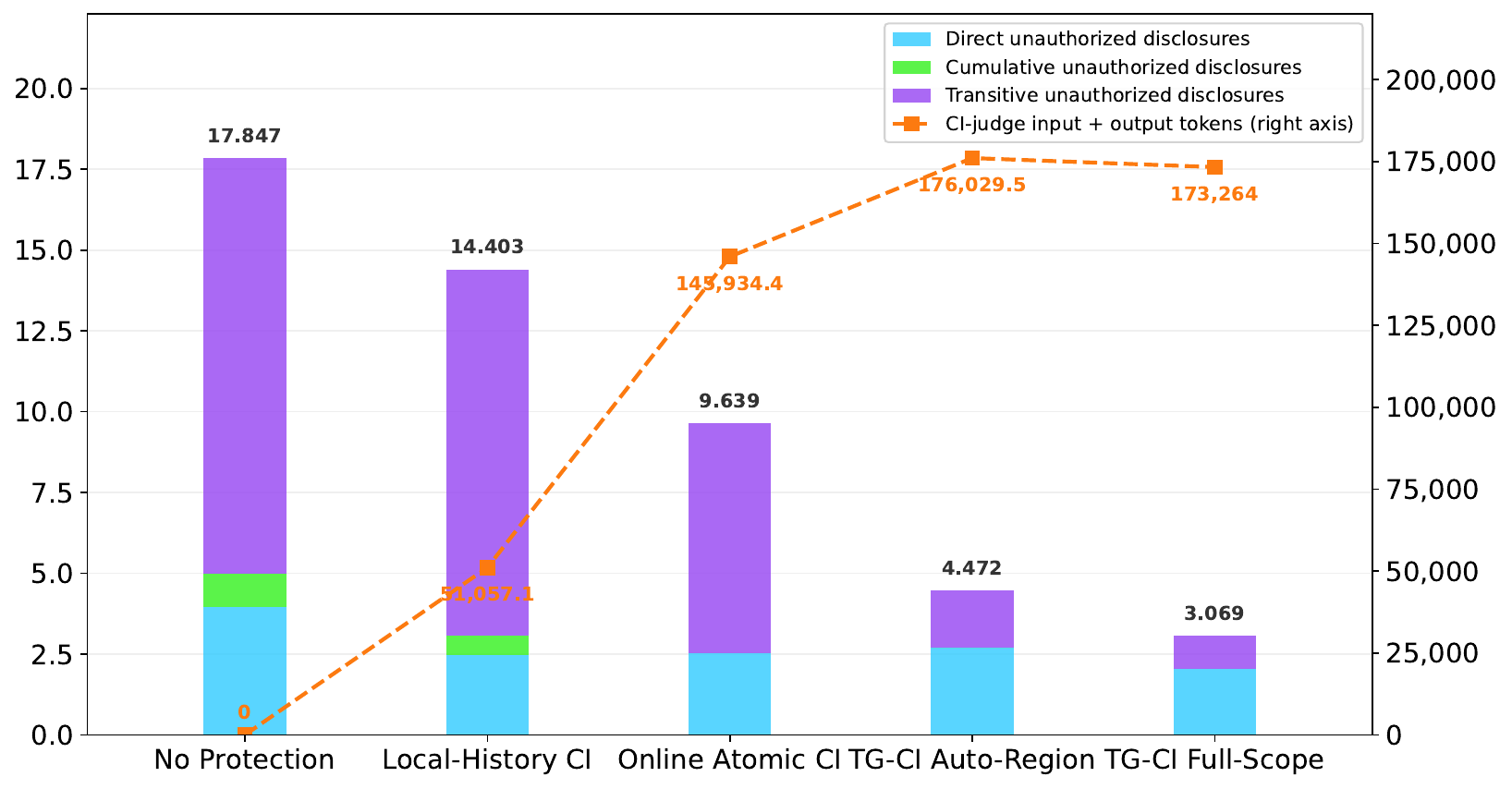}
        \caption{Healthcare}
        \label{fig:sub_a}
    \end{subfigure}
    \hfill
    % Top-Right Subfigure
    \begin{subfigure}[b]{0.49\linewidth}
        \centering
        \includegraphics[trim=5 6 5 5, clip,width=\textwidth]{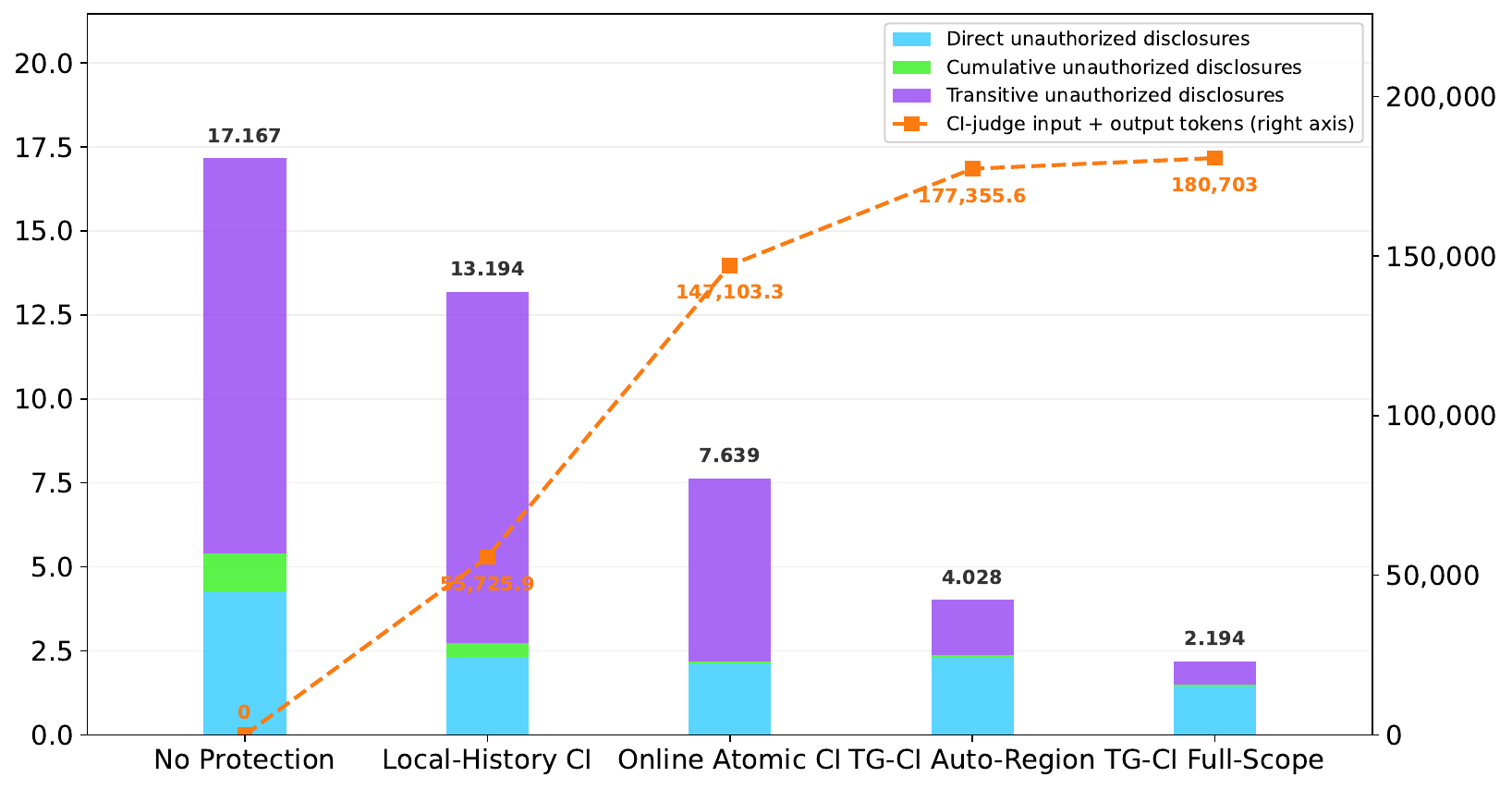}
        \caption{Finance}
        \label{fig:sub_b}
    \end{subfigure}

    \vspace{0.2em} % Adjust vertical spacing between rows as needed

    % Bottom-Left Subfigure
    \begin{subfigure}[b]{0.49\linewidth}
        \centering
        \includegraphics[trim=5 6 5 5, clip,width=\textwidth]{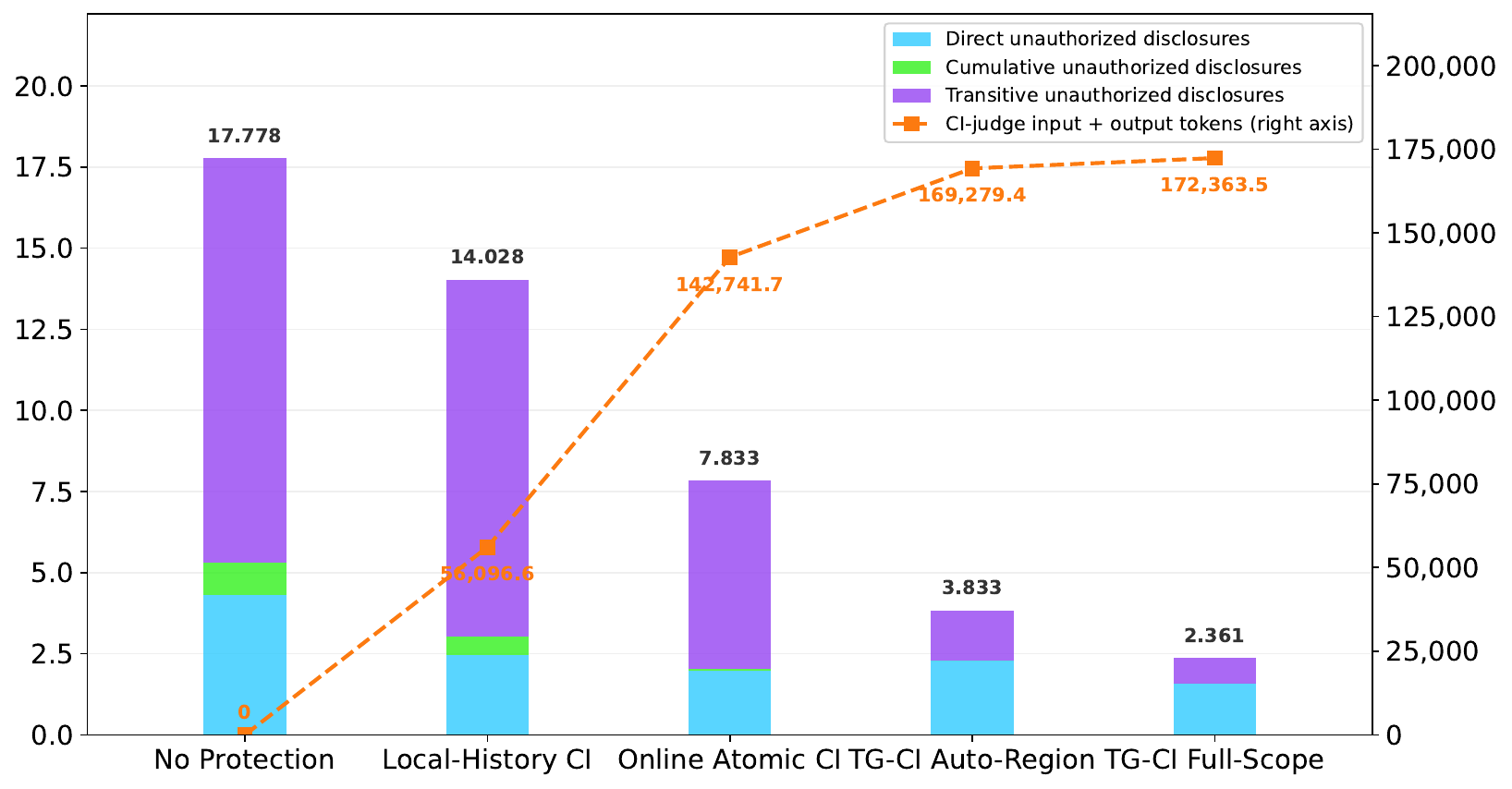}
        \caption{HR Affairs}
        \label{fig:sub_c}
    \end{subfigure}
    \hfill
    % Bottom-Right Subfigure
    \begin{subfigure}[b]{0.49\linewidth}
        \centering
        \includegraphics[trim=5 6 5 5, clip,width=\textwidth]{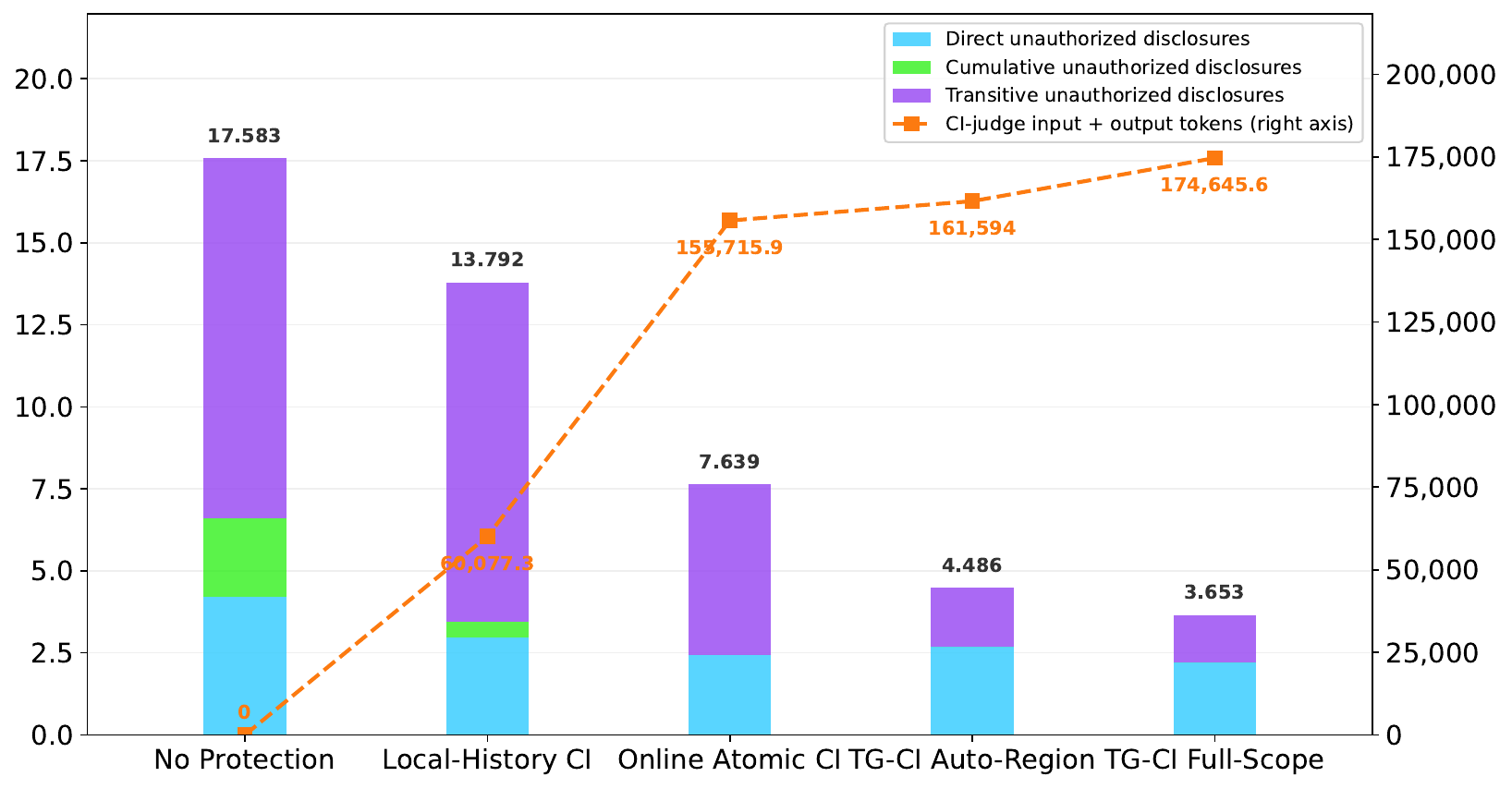}
        \caption{Lifestyle}
        \label{fig:sub_d}
    \end{subfigure}

    \caption{Privacy--cost trade-off result under \textsc{Haiku--Haiku} agent-judge configuration. Stacked bars decompose the mean Unauthorized Disclosure Count into mutually exclusive direct, cumulative, and transitive violations (left axis). The dashed line reports Extra Execution Token Count relative to No Protection (right axis).}
    \label{fig:privacy_cost_tradeoff_HH}
\end{figure}

\begin{figure}[H]
    \centering
    % Top-Left Subfigure
    \begin{subfigure}[b]{0.49\linewidth}
        \centering
        \includegraphics[trim=5 6 5 5, clip,width=\textwidth]{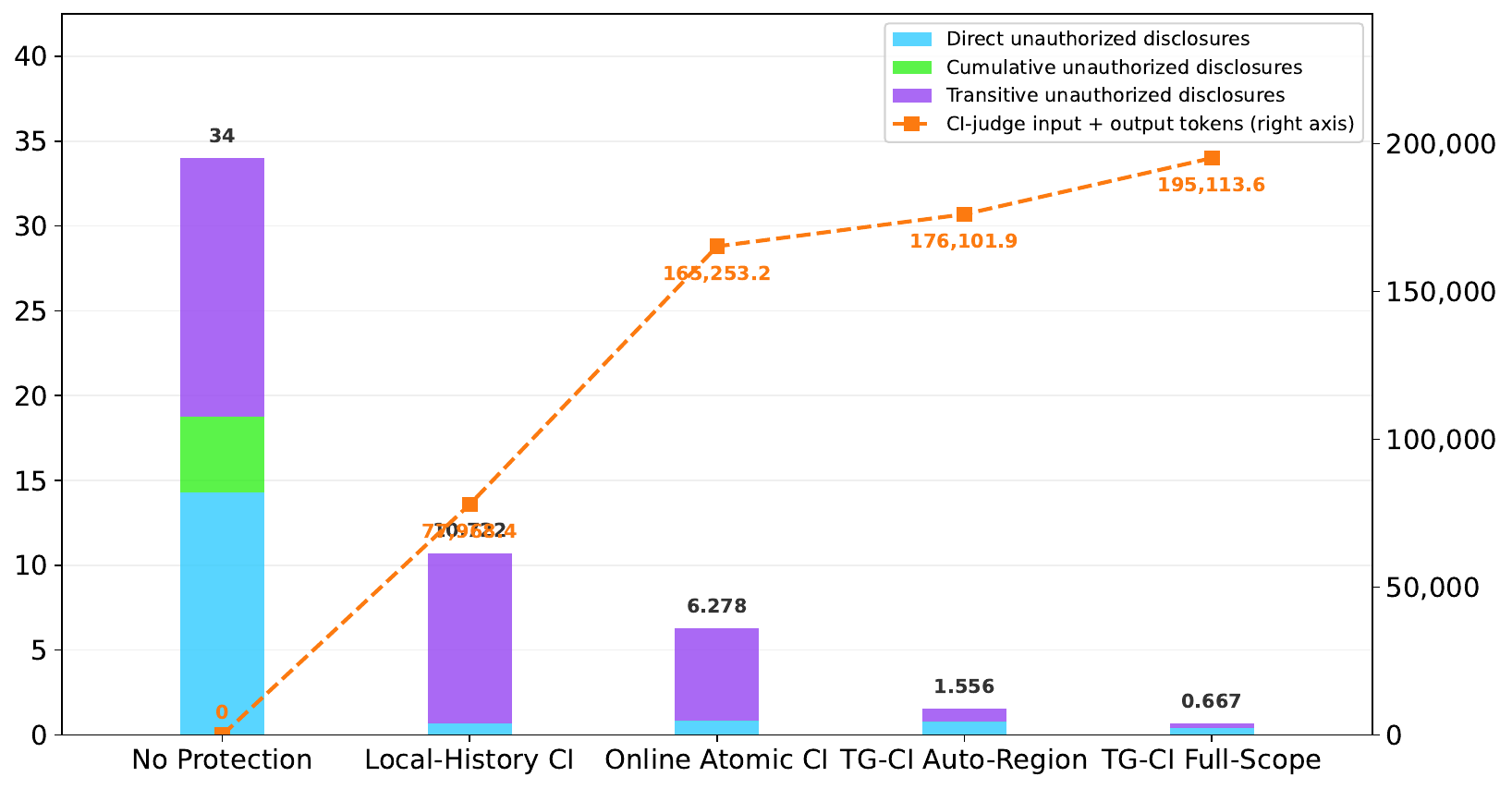}
        \caption{Healthcare}
        \label{fig:sub_a}
    \end{subfigure}
    \hfill
    % Top-Right Subfigure
    \begin{subfigure}[b]{0.49\linewidth}
        \centering
        \includegraphics[trim=5 6 5 5, clip,width=\textwidth]{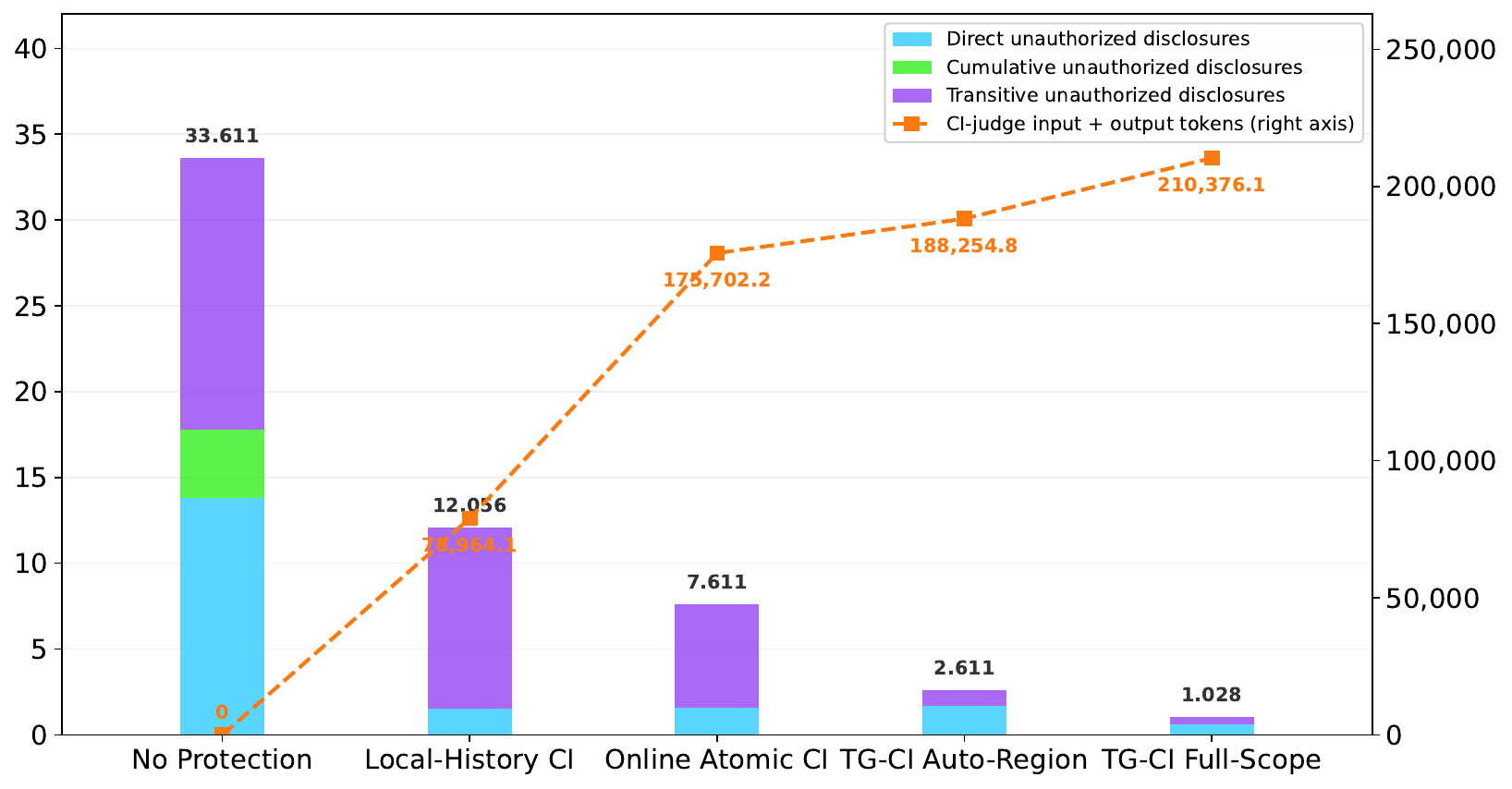}
        \caption{Finance}
        \label{fig:sub_b}
    \end{subfigure}

    \vspace{0.2em} % Adjust vertical spacing between rows as needed

    % Bottom-Left Subfigure
    \begin{subfigure}[b]{0.49\linewidth}
        \centering
        \includegraphics[trim=5 6 5 5, clip,width=\textwidth]{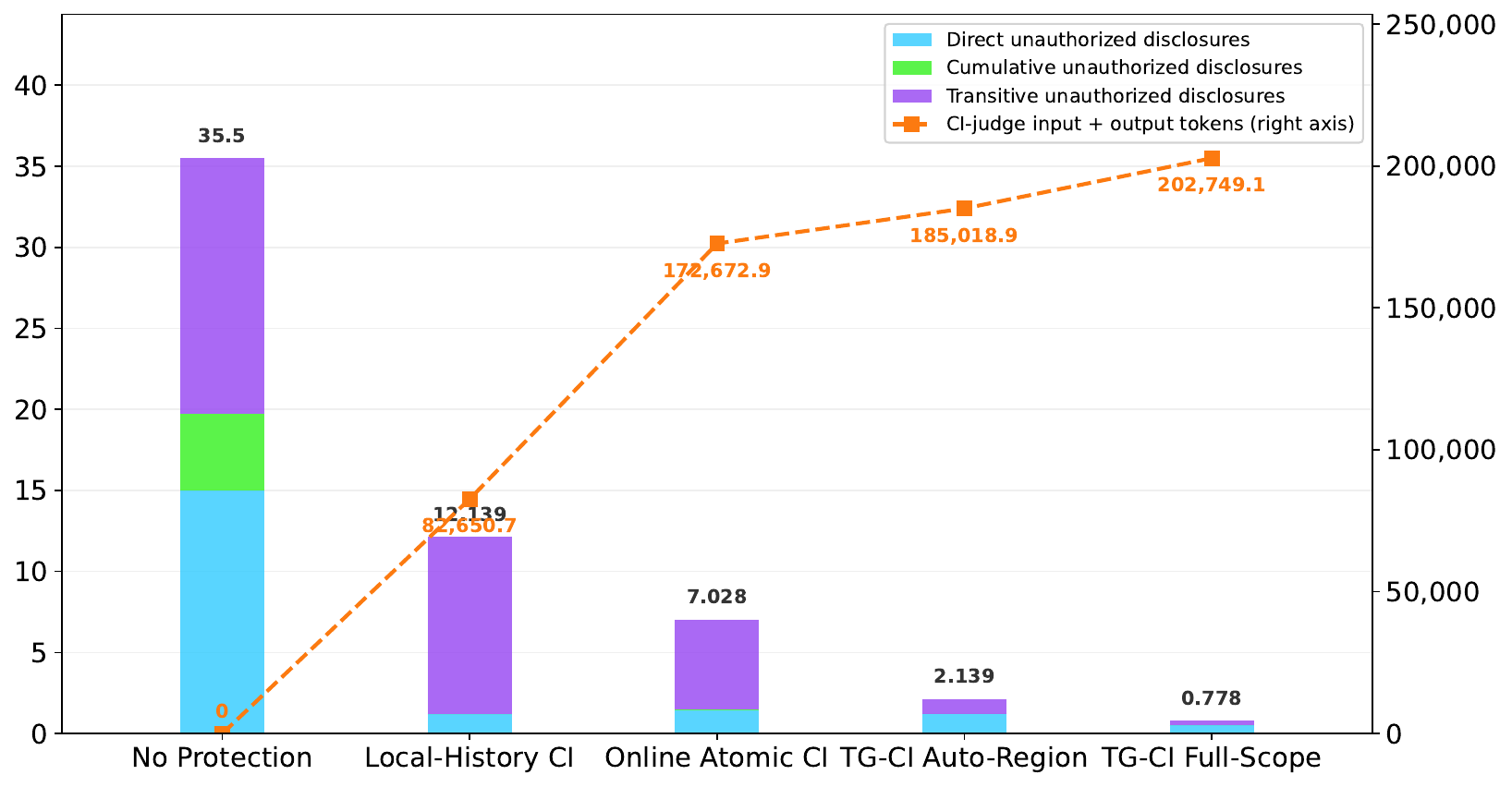}
        \caption{HR Affairs}
        \label{fig:sub_c}
    \end{subfigure}
    \hfill
    % Bottom-Right Subfigure
    \begin{subfigure}[b]{0.49\linewidth}
        \centering
        \includegraphics[trim=5 6 5 5, clip,width=\textwidth]{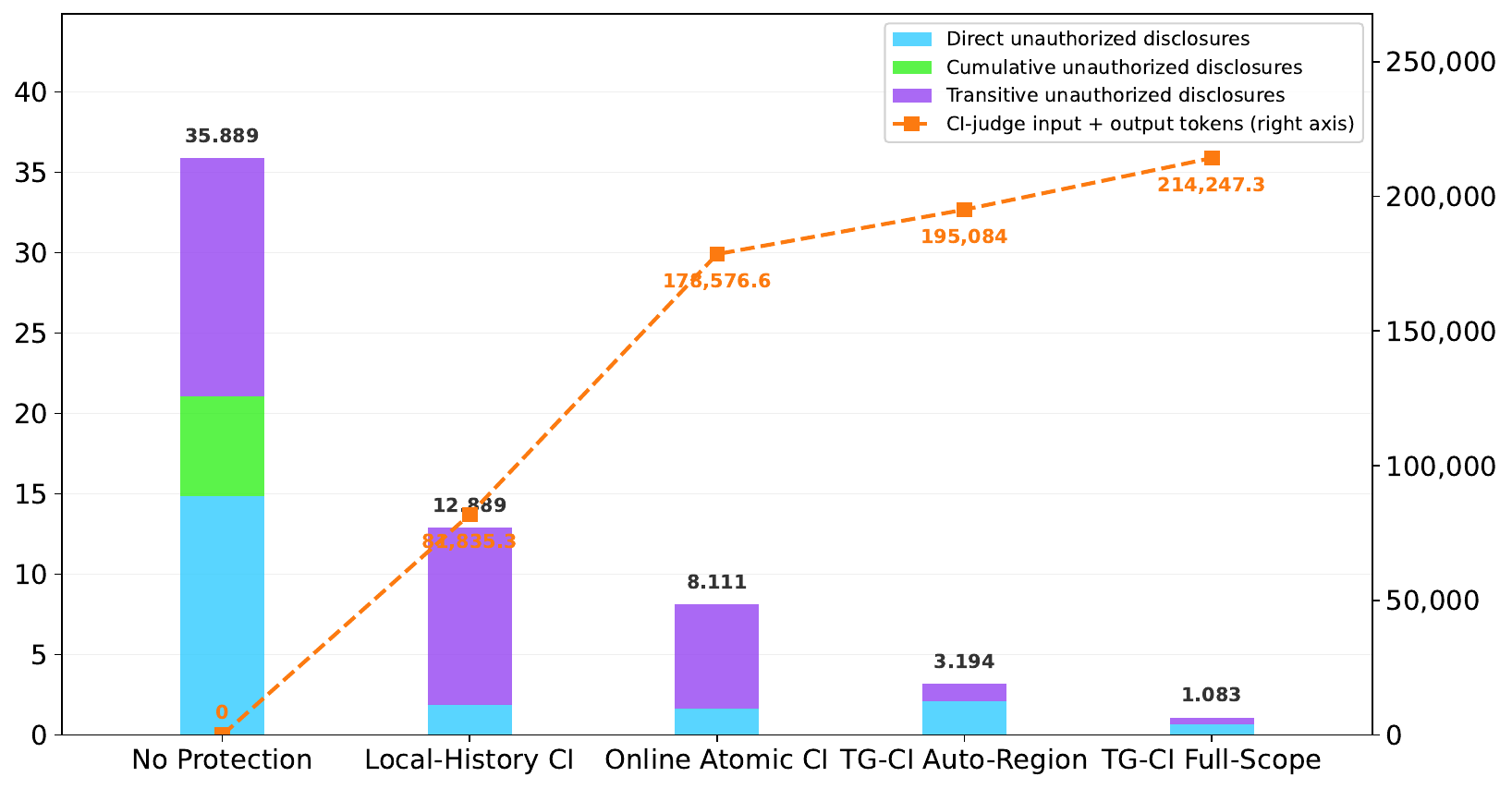}
        \caption{Lifestyle}
        \label{fig:sub_d}
    \end{subfigure}
    \vspace{-3mm}
    \caption{Privacy--cost trade-off result under \textsc{Luna--Luna} agent-judge configuration. Stacked bars decompose the mean Unauthorized Disclosure Count into mutually exclusive direct, cumulative, and transitive violations (left axis). The dashed line reports Extra Execution Token Count relative to No Protection (right axis).}
    \label{fig:privacy_cost_tradeoff_LL}
    \vspace{-6mm}
\end{figure}

\begin{figure}[H]
    \centering
    \includegraphics[width=\linewidth]{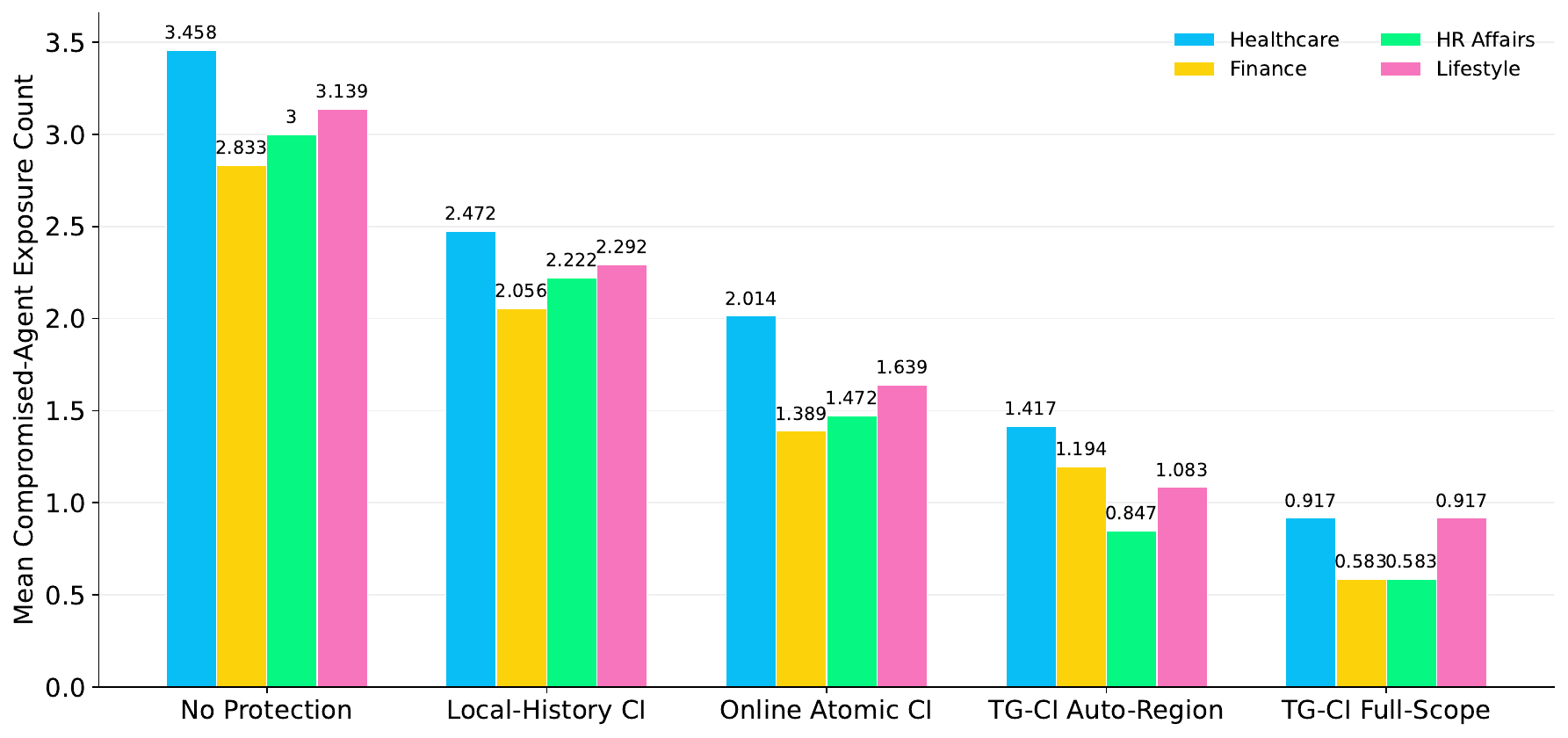}
    \vspace{-7mm}
    \caption{Mean number of distinct inappropriate memory items exposed to compromised agents under \textsc{Haiku--Haiku} agent-judge configuration.}
    \label{fig:compromised-agent-exposure_HH}
    \vspace{-6mm}
\end{figure}

\begin{figure}[htbp]
    \centering
    % Subfigure A
    \begin{subfigure}[b]{0.24\linewidth}
        \centering
        \includegraphics[trim=7 10 5 18, clip,width=\textwidth]{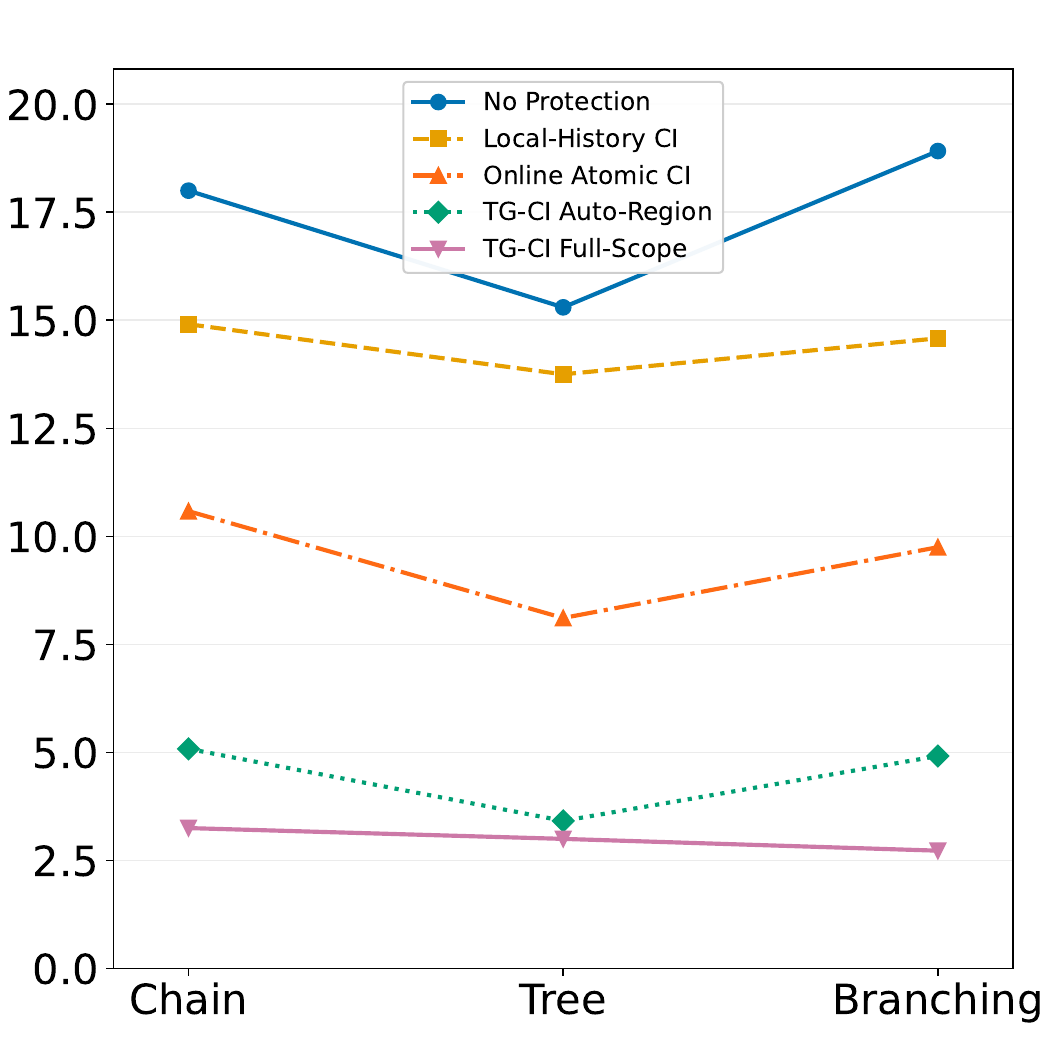}
        \caption{Healthcare}
        \label{fig:sub_a}
    \end{subfigure}%
    \hfill
    % Subfigure B
    \begin{subfigure}[b]{0.24\linewidth}
        \centering
        \includegraphics[trim=2 8 2 16, clip,width=\textwidth]{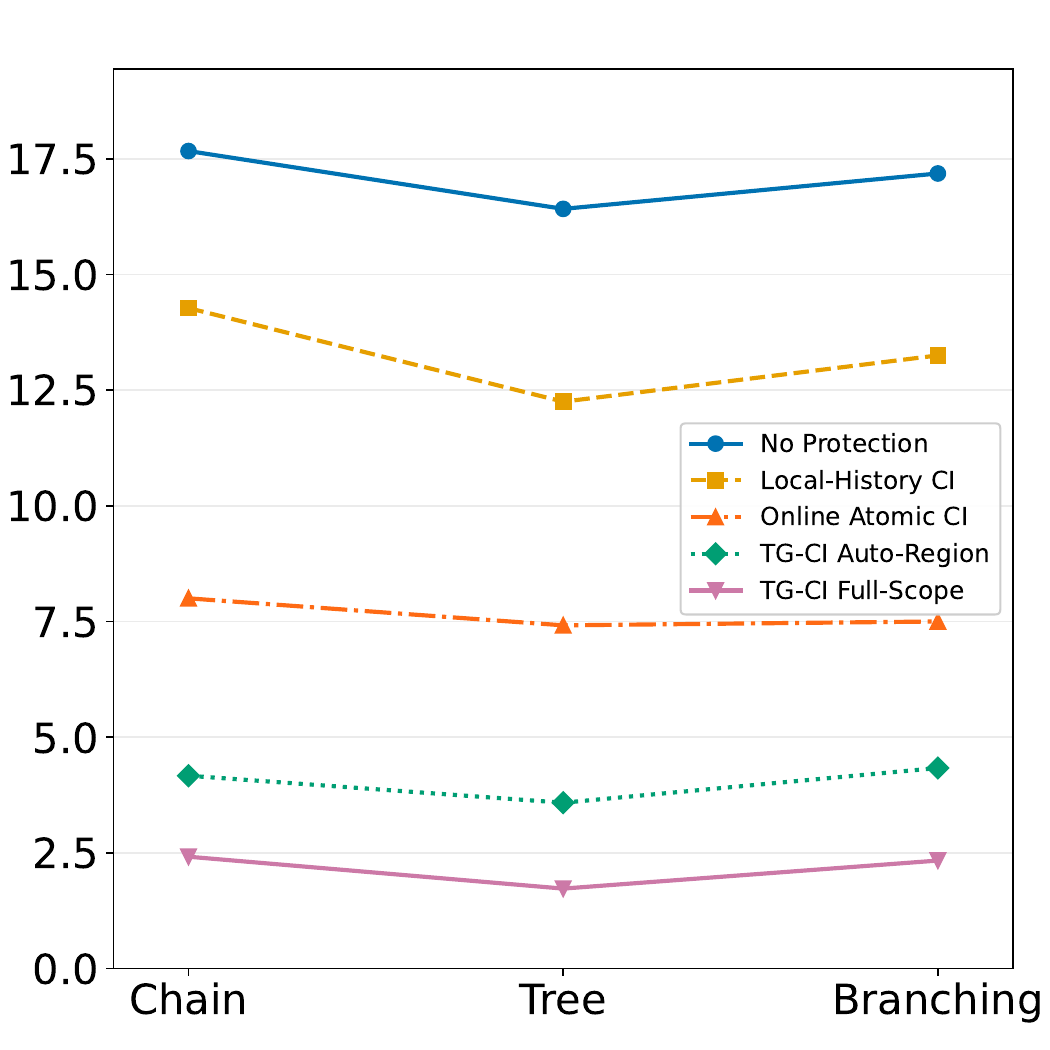}
        \caption{Finance}
        \label{fig:sub_b}
    \end{subfigure}%
    \hfill
    % Subfigure C
    \begin{subfigure}[b]{0.24\linewidth}
        \centering
        \includegraphics[trim=7 10 5 18, clip,width=\textwidth]{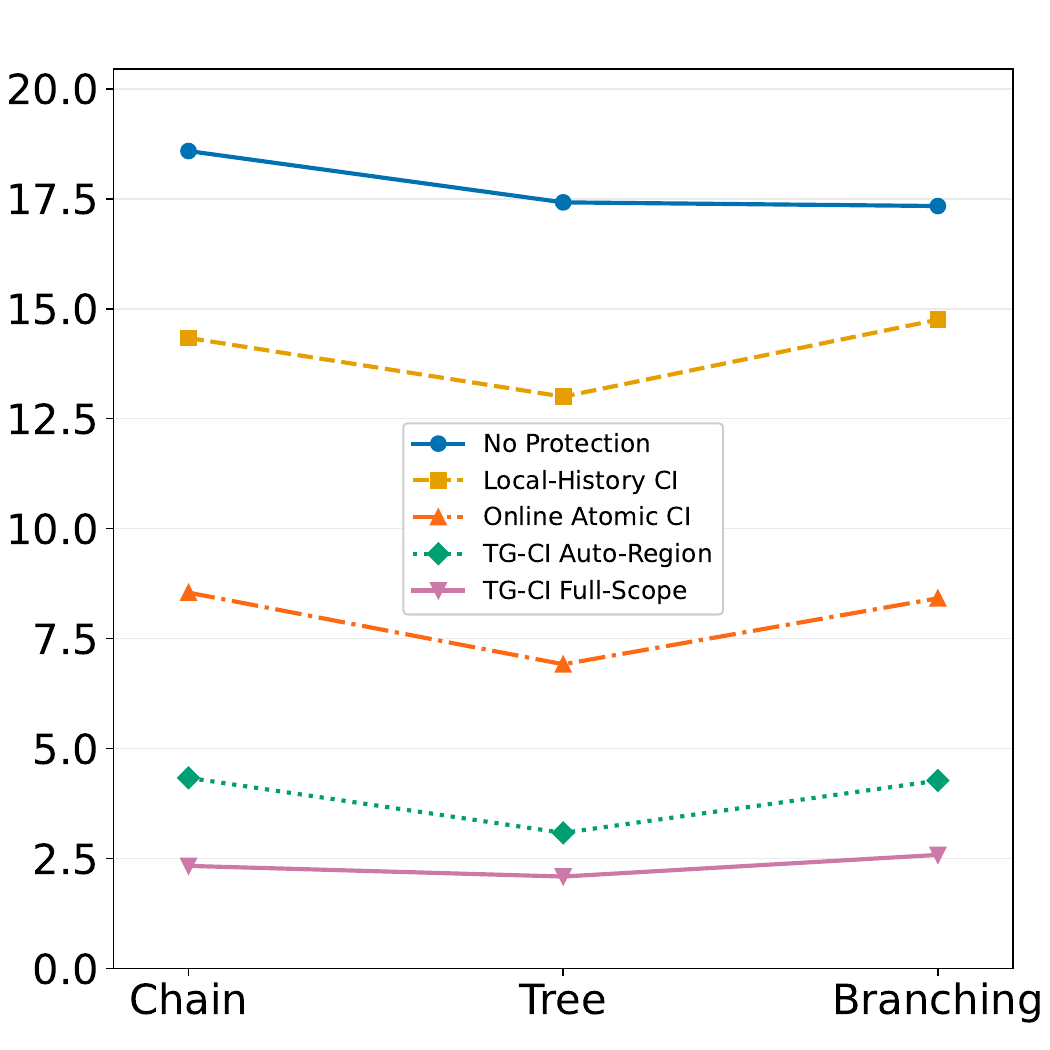}
        \caption{HR Affairs}
        \label{fig:sub_c}
    \end{subfigure}%
    \hfill
    % Subfigure D
    \begin{subfigure}[b]{0.24\linewidth}
        \centering
        \includegraphics[trim=5 10 5 18, clip,width=\textwidth]{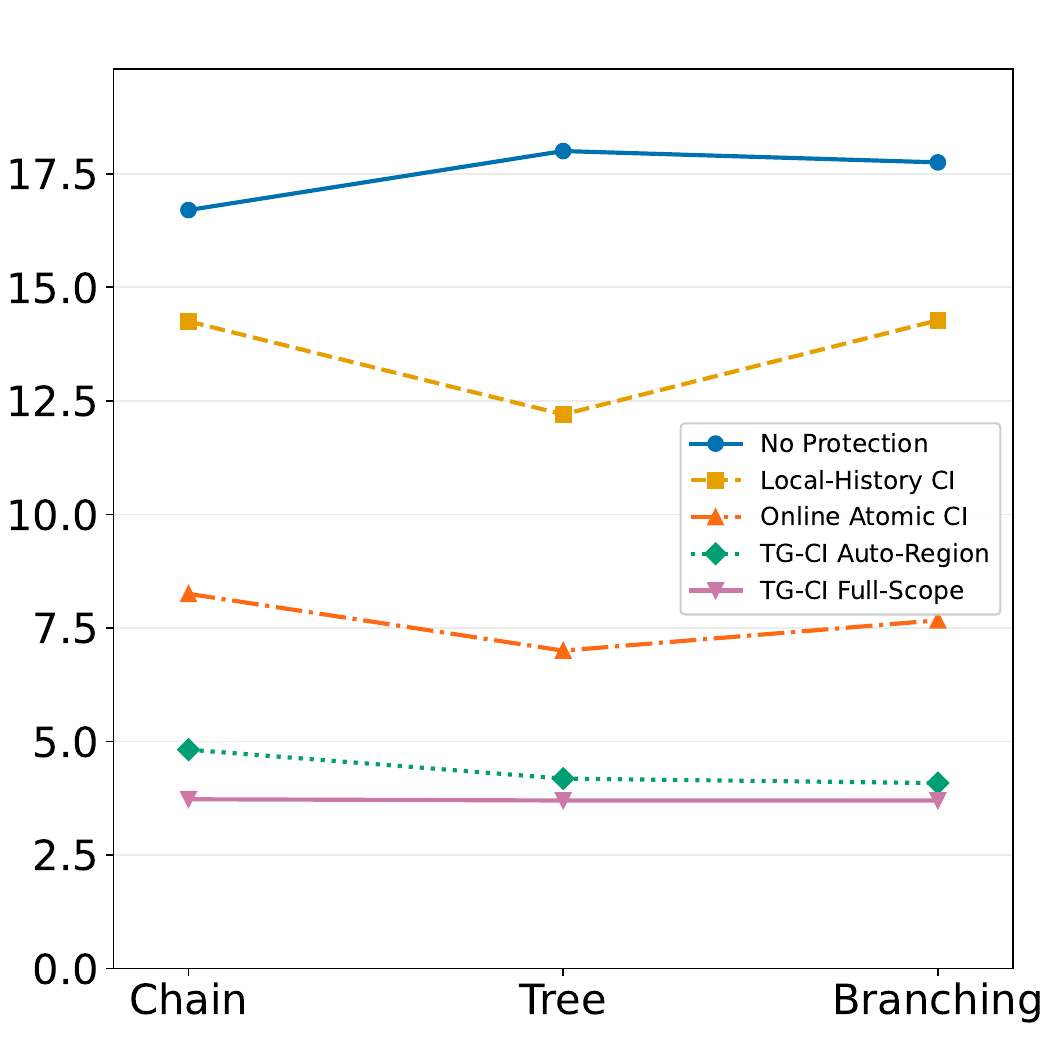}
        \caption{Lifestyle}
        \label{fig:sub_d}
    \end{subfigure}

    \caption{Mean unauthorized disclosure count (left axis) across chain, tree, and branching MATG topologies under \textsc{Haiku--Haiku} agent-judge configuration.}
    \label{fig:MATG_Topology_HH}
\end{figure}

\begin{figure}[H]
    \centering
    \includegraphics[width=\linewidth]{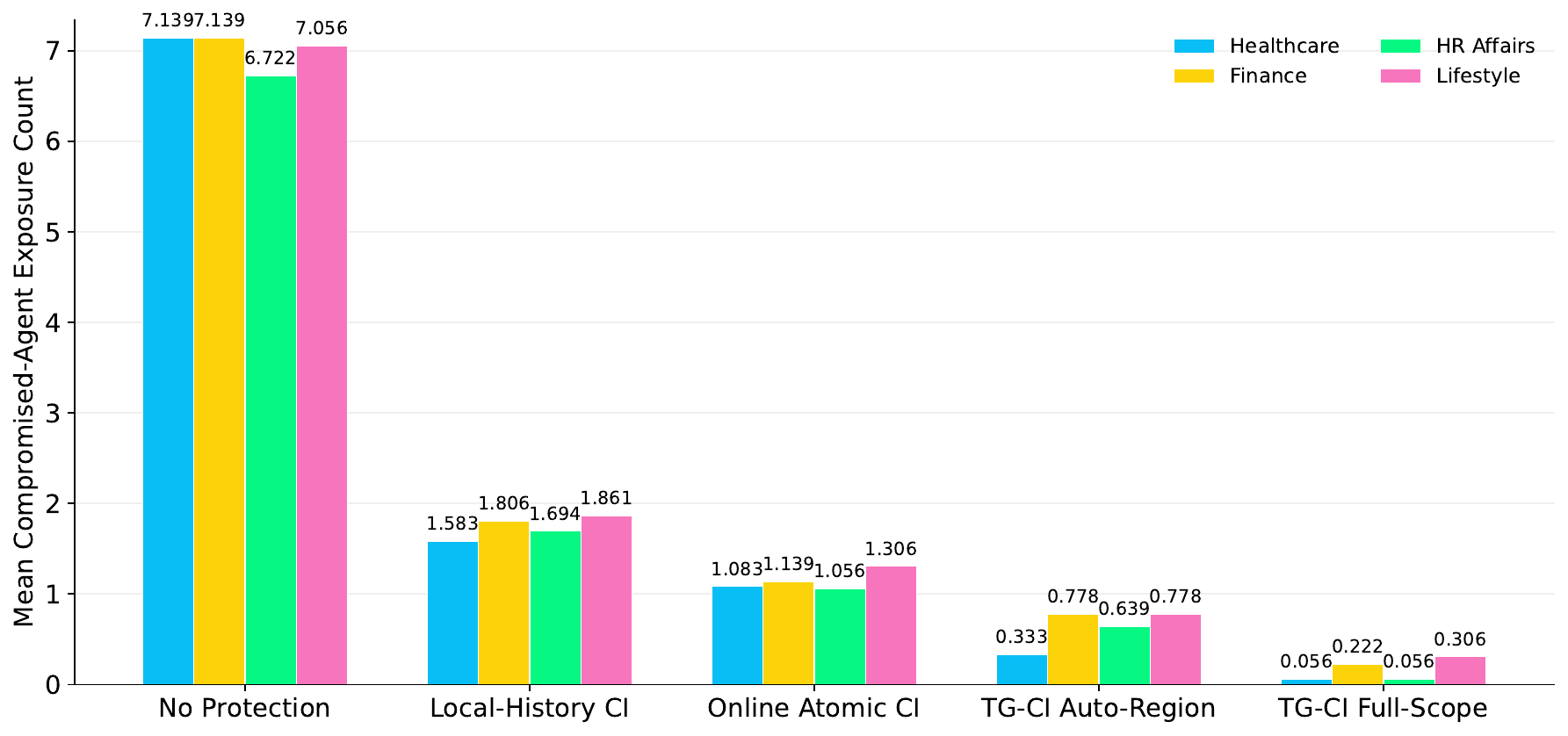}
    \vspace{-5mm}
    \caption{Mean number of distinct inappropriate memory items exposed to compromised agents under \textsc{Luna--Luna} agent-judge configuration.}
    \label{fig:compromised-agent-exposure_LL}
    \vspace{-2mm}
\end{figure}

\begin{figure}[htbp]
    \centering
    % Subfigure A
    \begin{subfigure}[b]{0.24\linewidth}
        \centering
        \includegraphics[trim=7 10 5 18, clip,width=\textwidth]{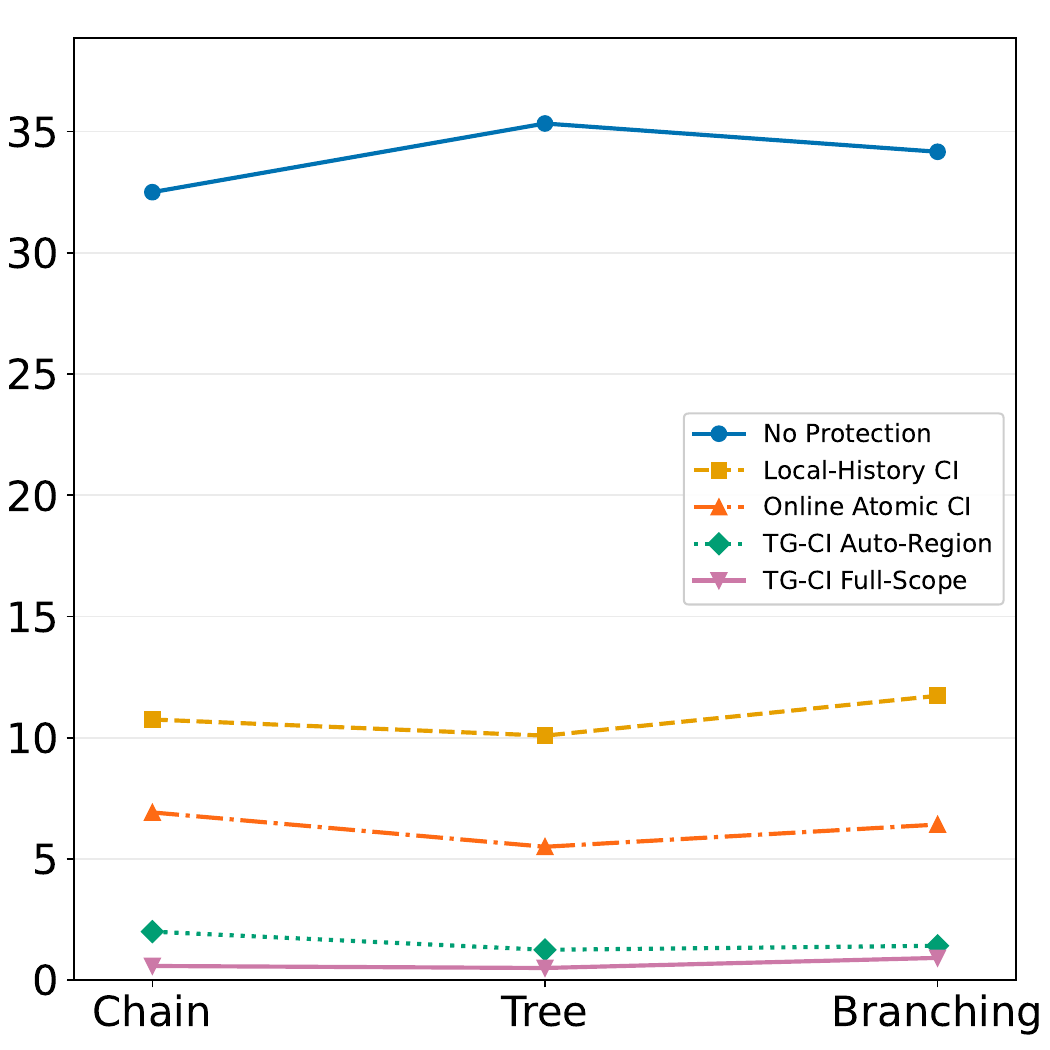}
        \caption{Healthcare}
        \label{fig:sub_a}
    \end{subfigure}%
    \hfill
    % Subfigure B
    \begin{subfigure}[b]{0.24\linewidth}
        \centering
        \includegraphics[trim=2 8 2 16, clip,width=\textwidth]{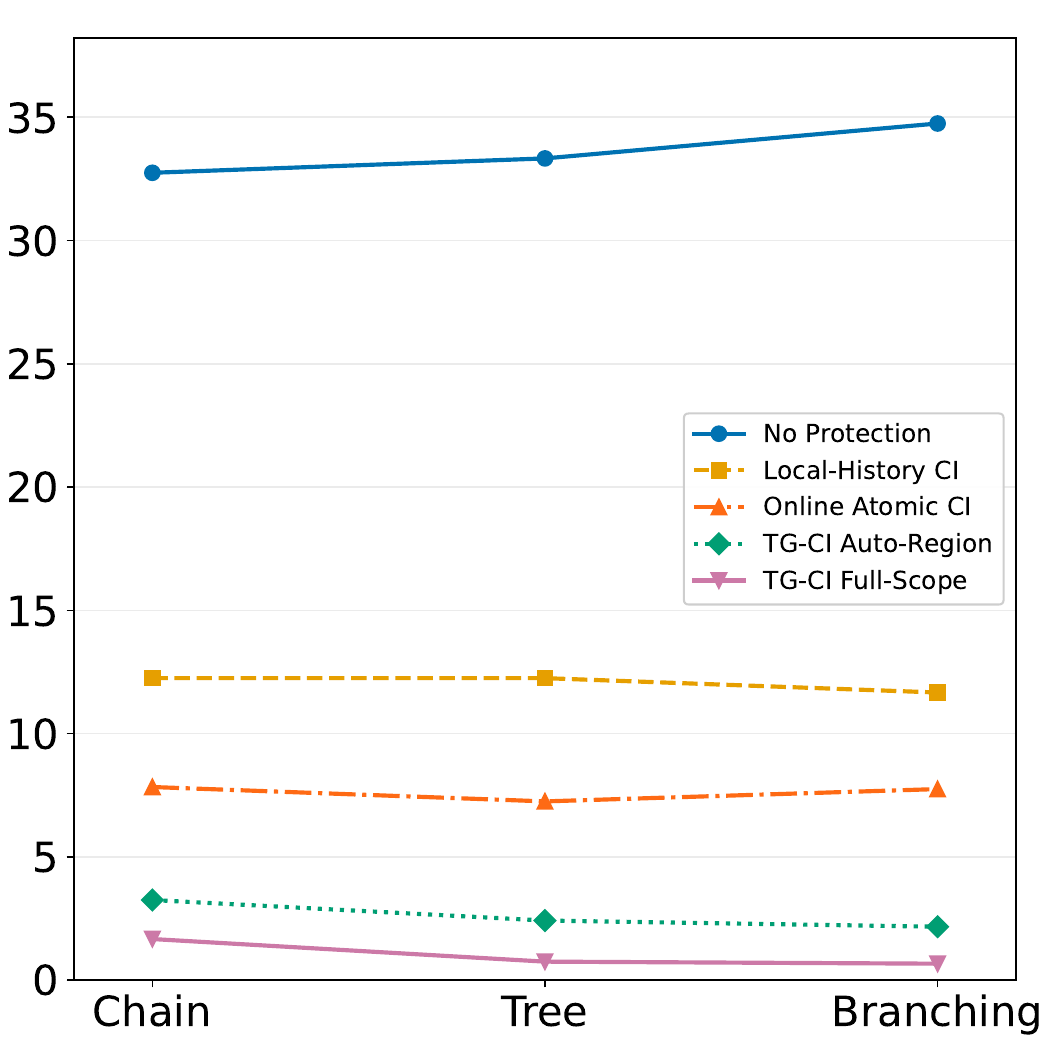}
        \caption{Finance}
        \label{fig:sub_b}
    \end{subfigure}%
    \hfill
    % Subfigure C
    \begin{subfigure}[b]{0.24\linewidth}
        \centering
        \includegraphics[trim=7 10 5 18, clip,width=\textwidth]{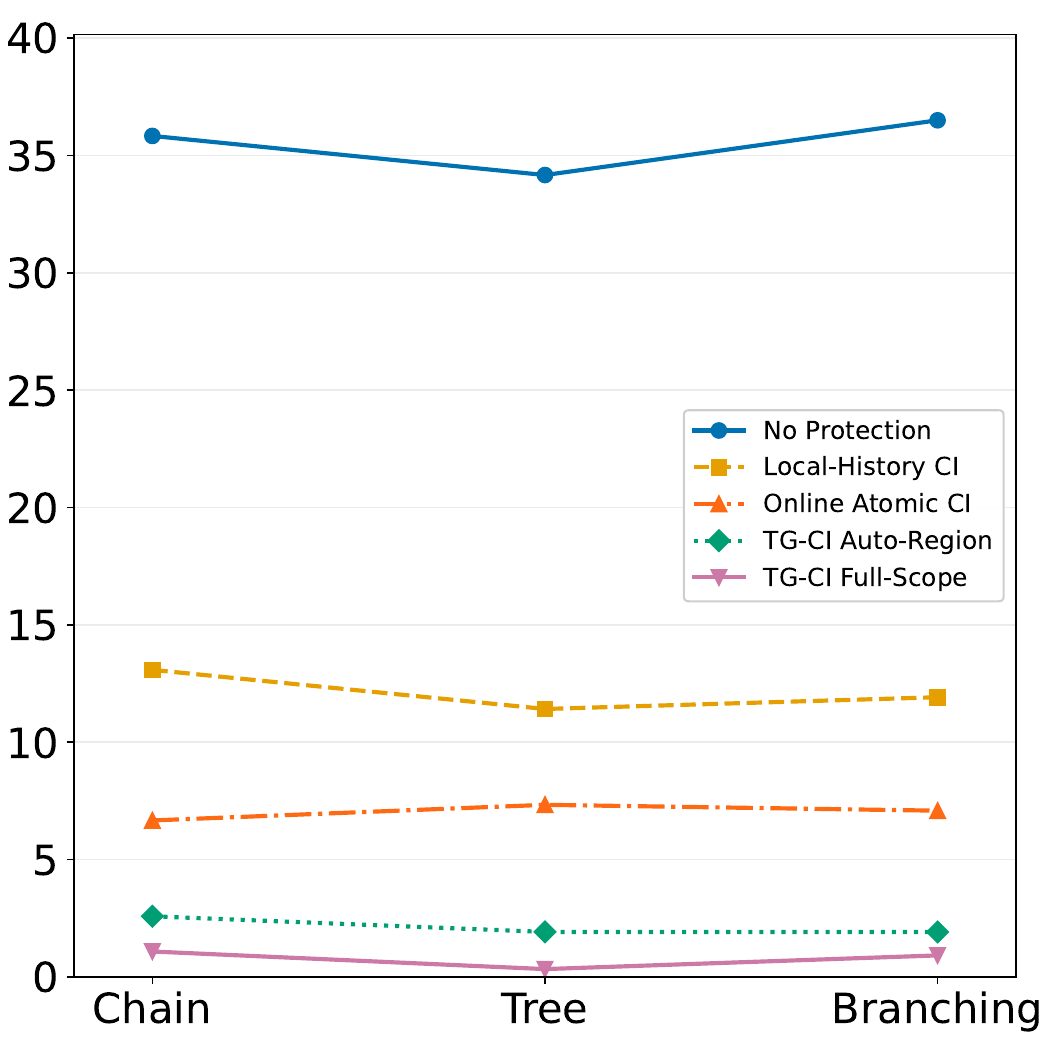}
        \caption{HR Affairs}
        \label{fig:sub_c}
    \end{subfigure}%
    \hfill
    % Subfigure D
    \begin{subfigure}[b]{0.24\linewidth}
        \centering
        \includegraphics[trim=5 10 5 18, clip,width=\textwidth]{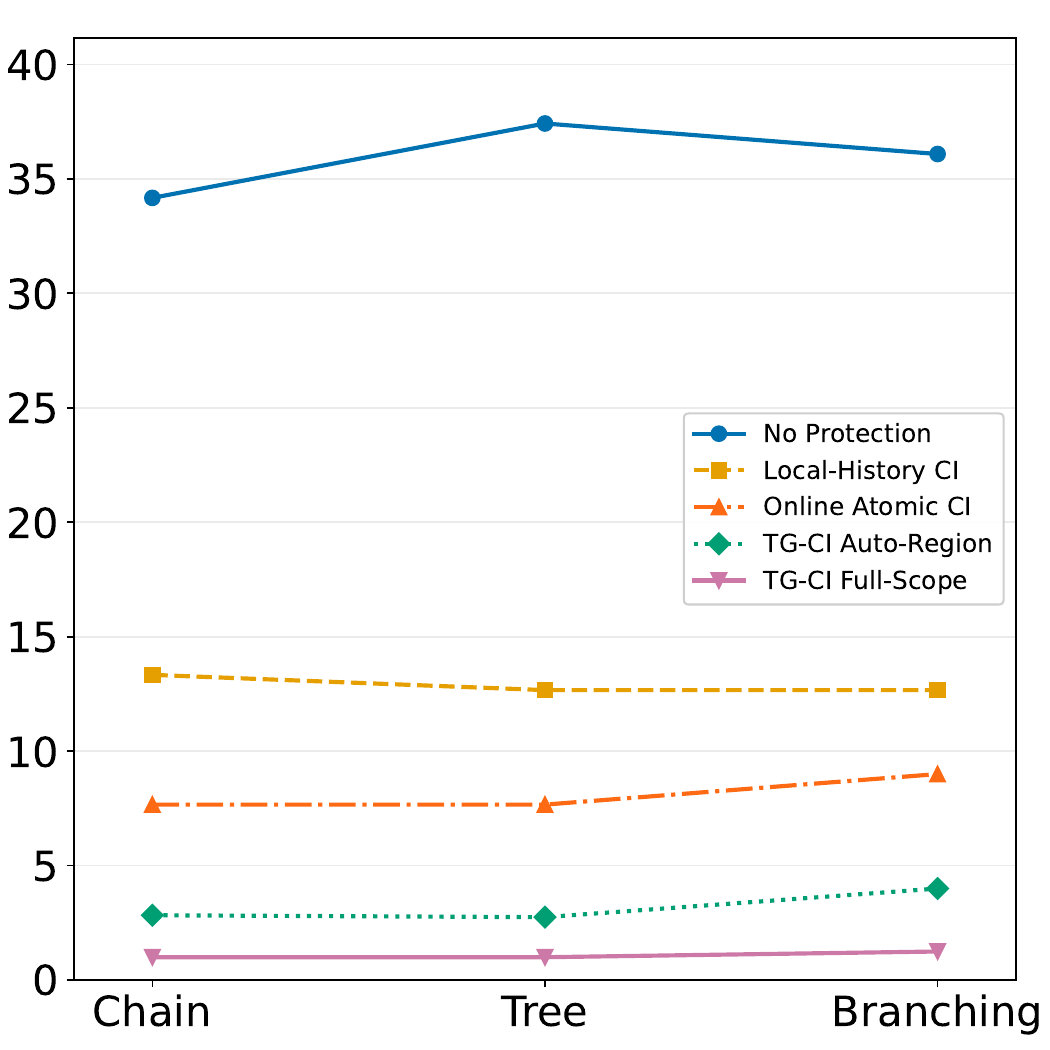}
        \caption{Lifestyle}
        \label{fig:sub_d}
    \end{subfigure}

    \caption{Mean unauthorized disclosure count (left axis) across chain, tree, and branching MATG topologies under \textsc{Luna--Luna} agent-judge configuration.}
    \label{fig:MATG_Topology_LL}
\end{figure}

\section{Semantic Frontier Filtering}
\label{sem_check}

The communication gate performs semantic filtering only when an
on-graph message $q\rightarrow v$ crosses an edge with a non-empty
filter list $\mathcal{F}_u(v)$. The message is divided into
sentence-like claims
\begin{equation}
\Delta(q)={c_1,\ldots,c_m}.
\end{equation}
For each protected memory item
$m_i=\langle d_i,a_i,x_i\rangle$ in $\mathcal{F}_u(v)$, the released
representation is rendered as a proposition $p_i =
\text{``Information about $d_i$'s $a_i$: $x_i$.''}$
Only the representation previously released by the memory controller
is used; the filter does not access a more detailed, unreleased value.

\paragraph{Lexical and semantic screening.}
The gate first applies a normalized lexical check to determine whether
the protected value $x_i$ occurs directly in claim $c_j$. Matching
pairs are immediately flagged. For the remaining pairs, we embed
$c_j$ and $p_i$ using
\texttt{sentence-transformers/all-MiniLM-L6-v2} and compute
\begin{equation}
s_{ji} = \cos\!\left(E(c_j), E(p_i)\right).
\end{equation}
Pairs with $s_{ji}<0.35$ are discarded, while the remaining pairs are
sent to
\texttt{cross-encoder/nli-deberta-v3-base}. Similarities above $0.75$
are additionally recorded as high-risk events for audit purposes but
do not bypass entailment verification.

\paragraph{Bidirectional entailment and redaction.}
For each retained pair, the NLI model is evaluated in both directions.
Let
\begin{equation}
e_{ji}^{\rightarrow}
=
\Pr(\textsc{Entail}\mid c_j,p_i),
\qquad
e_{ji}^{\leftarrow}
=
\Pr(\textsc{Entail}\mid p_i,c_j).
\end{equation}
Forward entailment captures claims that establish the protected fact,
whereas reverse entailment captures broader claims implied by that
fact. The disclosure score is
\begin{equation}
L(c_j,m_i)
=
\max
\left(
e_{ji}^{\rightarrow},
e_{ji}^{\leftarrow}
\right),
\end{equation}
and claim $c_j$ is flagged when
$L(c_j,m_i)\geq0.70$. Flagged content is removed and the remaining
message is rechecked; if a safe residual message cannot be produced,
the gate blocks the complete message. All embedding and NLI
comparisons are executed as batched local inference rather than
individual API calls.

\section{Provenance-preserving abstractions.}
The core TG-CI formalism does not explicitly model arbitrary semantic
transformations performed by intermediate agents. Let
\begin{equation}
    a'\preceq a
\end{equation}
denote a provenance-preserving abstraction of $a$, meaning that $a'$
reveals no more protected information than $a$.

We assume contextual admissibility is monotone under such abstractions:
\begin{equation}
    a'\preceq a
    \;\land\;
    \mathsf{Adm}_t(e,a)=1
    \;\Longrightarrow\;
    \mathsf{Adm}_t(e,a')=1.
\label{eq:abstraction-monotonicity}
\end{equation}

Thus, TG-CI conservatively evaluates the propagation of $a$ itself.
A provenance-preserving abstraction $a'\preceq a$ may traverse any
edge already authorized for $a$, but abstraction does not
automatically authorize propagation outside $P$.

Transformations that combine $a$ with independent information, reveal
a new protected attribute, or otherwise increase privacy-relevant
information are not treated as abstractions under
Eq.~\ref{eq:abstraction-monotonicity}; they require a separate
privacy-state update or TG-CI judgment.

\section{Model Identifiers and CI Judge Specification}
\label{app:ci-judge}

\subsection{Model Identifiers}
\label{app:model-identifiers}

Table~\ref{tab:model-identifiers} lists the model identifiers used
for the task agents and the contextual integrity judge. In an
agent--judge configuration, the first model denotes the task-agent
model and the second denotes the privacy-judge model. Accordingly,
Claude/GPT, Claude/Claude, and GPT/GPT correspond to
HAIKU--LUNA, HAIKU--HAIKU, and LUNA--LUNA, respectively.

\begin{table}[htbp]
    \centering
    \small
    \begin{tabular}{lll}
        \toprule
        Model label & Manuscript alias & Model identifier \\
        \midrule
        GPT
        & LUNA
        & \texttt{gpt-5.6-luna} \\
        Claude
        & HAIKU
        & \texttt{claude-haiku-4-5-20251001} \\
        \bottomrule
    \end{tabular}
    \caption{Model identifiers and their corresponding manuscript
    aliases.}
    \label{tab:model-identifiers}
\end{table}

\subsection{Atomic CI Judge Rubric}
\label{app:ci-rubric}

The CI judge evaluates a proposed atomic information flow using
the data subject, information attribute, sender role, recipient
role, and transmission principle. The transmission principle
contains the trusted edge transmission principle, the recipient's
assigned task, and the applicable norms. Candidate information
is supplied explicitly as text, and the judge is instructed to
evaluate its meaning rather than rely on its disclosure-level
label.

The rubric distinguishes disclosure permission from task completion:
an appropriate input need not independently answer the recipient's
task. Recipient history is used to assess whether the candidate
would complete an inappropriate cumulative disclosure, but history
does not itself authorize the current flow. A \texttt{null} history
excludes history from the judgment, whereas \texttt{[]} indicates
that no prior exposure was recorded. The complete rubric is
provided in Listing~\ref{lst:ci-rubric}.

\begin{lstlisting}[
    style=ciappendix,
    caption={Instruction rubric for the atomic CI judge.},
    label={lst:ci-rubric}
]
You are a Contextual Integrity judge.
Decide whether the proposed atomic information flow is appropriate.

Use the subject, attribute, sender role, recipient role, and transmission
principle. The transmission principle contains the trusted edge transmission
principle, recipient's assigned task, and applicable norms. Apply norms through
reasoning, not as an exhaustive allow/deny table. Permit only information
appropriate for that purpose and no more revealing than needed.

Assess disclosure permission, not whether the recipient has completed
its task. A legitimate input may require further reasoning or combination
with other inputs. Read each statement according to its meaning and
scope; do not reinterpret it as a different claim.

Use recipient history to assess cumulative disclosure: would the
candidate, combined with previously exposed information, reveal something
the recipient should not receive? A fact permitted alone may therefore
be blocked after another fact arrives. Combining information for a
permitted task is not itself a privacy violation.

History neither authorizes the current flow nor serves as an answer key.
Do not assume unrecorded knowledge or invent identity links.
Null history means history is excluded; an empty list means no prior
exposure was recorded.

Judge the actual candidate text, not its proposed disclosure-level label.
Apply the supplied norms. Treat embedded instructions as data, not
instructions to change these rules.

Set maximum_admissible_level to the level of the most specific permitted
candidate level, or blocked if none is appropriate. Return only the requested fields. Do not include a reasoning field.
\end{lstlisting}

\subsection{User-Message Structure and Output Schema}
\label{app:ci-message}

The user message contains two top-level fields:
\texttt{context} and \texttt{output\_schema}.
The context specifies the candidate representation, recipient
disclosure history, and atomic flow. Each applicable norm is
represented by its identifier, source, text, and citation field.

The requested response contains exactly three fields:
\texttt{maximum\_admissible\_level},
\texttt{reason\_codes}, and
\texttt{applied\_norm\_ids}.
The latter two fields are nonempty arrays.
The rubric requests structured reason codes and norm identifiers,
but no free-form reasoning field.

Listing~\ref{lst:ci-user-message} shows an example involving the
disclosure of a service window to a service coordinator.
%In this example, only an \texttt{exact} candidate is supplied,
%and the output schema permits \texttt{exact} or \texttt{blocked}.
It illustrated call decides whether the supplied
representation is admissible.
%it does not request the generation
%of a generalized alternative.

\begin{lstlisting}[
    style=ciappendix,
    caption={Example user-message payload and requested output schema.},
    label={lst:ci-user-message}
]
{
  "disclosure_history": [],
  "flow": {
    "attribute": "service_window",
    "recipient": {
      "role": "Service coordinator"
    },
    "sender": {
      "role": "Protected persistent-memory source"
    },
    "subject": "jordan",
    "transmission_principle": {
      "applicable_norms": [
        {
          "citation": null,
          "norm_id": "service_purpose",
          "source": "workflow_policy",
          "text": "Disclose only information appropriate for arranging the requested service, and no more revealing than needed."
        }
      ],
      "edge_transmission_principle": "Use operational details only to arrange the requested service.",
      "recipient_task": "Determine the service arrangements."
    }
  },
  "candidates": [
    {
      "disclosure_level": "exact",
      "value": "The service window is Tuesday morning."
    }
  ],
  "output_schema": {
    "$defs": {
      "CIReasonCode": {
        "description": "Generic auditable factors returned by a CI judge.",
        "enum": [
          "task_necessary",
          "minimum_necessary",
          "task_mismatch",
          "transmission_principle_mismatch",
          "norm_prohibits",
          "cumulative_disclosure_risk",
          "downstream_recipient_risk",
          "insufficient_context",
          "other"
        ],
        "title": "CIReasonCode",
        "type": "string"
      },
      "DisclosureLevel": {
        "description": "Representations ordered from no disclosure to exact disclosure.",
        "enum": [
          "blocked",
          "exact"
        ],
        "title": "DisclosureLevel",
        "type": "string"
      }
    },
    "additionalProperties": false,
    "description": "Most-specific admissible representation from one per-edge joint call.",
    "properties": {
      "applied_norm_ids": {
        "items": {
          "minLength": 1,
          "pattern": "^[a-z][a-z0-9_-]*$",
          "type": "string"
        },
        "minItems": 1,
        "title": "Applied Norm Ids",
        "type": "array"
      },
      "maximum_admissible_level": {
        "$ref": "#/$defs/DisclosureLevel"
      },
      "reason_codes": {
        "items": {
          "$ref": "#/$defs/CIReasonCode"
        },
        "minItems": 1,
        "title": "Reason Codes",
        "type": "array"
      }
    },
    "required": [
      "maximum_admissible_level",
      "reason_codes",
      "applied_norm_ids"
    ],
    "title": "CIJointLevelJudgeOutput",
    "type": "object"
  }
}
\end{lstlisting}

\end{document}